\newcommand{\oivs}{O\,\textsc{iv}]}
\newcommand{\oii}{[O\,\textsc{ii}]}
\newcommand{\oiii}{[O\,\textsc{iii}]}
\newcommand{\nii}{[N\,\textsc{ii}]}
\newcommand{\feii}{[Fe\,\textsc{ii}]}
\newcommand{\feiii}{[Fe\,\textsc{iii}]}
\newcommand{\feiv}{[Fe\,\textsc{iv}]}
\newcommand{\fev}{[Fe\,\textsc{v}]}
\newcommand{\cliii}{[Cl\,\textsc{iii}]}
\newcommand{\siii}{[S\,\textsc{iii}]}
\newcommand{\sii}{[S\,\textsc{ii}]}
\newcommand{\ariv}{[Ar\,\textsc{iv}]}
\newcommand{\hi}{H\,\textsc{i}}
\newcommand{\hii}{H\,\textsc{ii}}
\newcommand{\heii}{He\,\textsc{ii}}
\newcommand{\hei}{He\,\textsc{i}}
\newcommand{\ariii}{[Ar\,\textsc{iii}]}
\newcommand{\oiirls}{O\,\textsc{ii}}

\documentclass{aa}  
\usepackage{multirow}
\usepackage{soul}
\usepackage{graphicx}
\usepackage{placeins}

\usepackage{txfonts}
\usepackage{xcolor}

\usepackage[colorlinks=true,linkcolor=blue,citecolor=blue,urlcolor=blue]{hyperref}

\begin{document}

   \title{Gas-phase Fe/O and Fe/N abundances in Star-Forming Regions}

   \subtitle{Relations between nucleosynthesis, metallicity and dust}

   \author{ J. E. M\'endez-Delgado
          \inst{1},
          K. Kreckel\inst{1},
          C. Esteban\inst{2,3},
          J. Garc\'{\i}a-Rojas\inst{2,3}, L. Carigi\inst{4},
          A.A.C.\ Sander\inst{1},
          M. Palla\inst{5,6},
          M. Chru{\'s}li{\'n}ska \inst{7},
          I. De Looze\inst{8},
          M. Rela{\~n}o\inst{9,10},
          S.A. van der Giessen \inst{8,9},
          E. Reyes-Rodr\'{\i}guez\inst{2,3},
          S. F. S\'anchez\inst{11,2}
     }

   \institute{Astronomisches Rechen-Institut, Zentrum f\"ur Astronomie der Universit\"at Heidelberg, Mönchhofstraße 12-14, D-69120 Heidelberg, Germany \email{jemd@uni-heidelberg.de} 
        \and
        Instituto de Astrof\'isica de Canarias, E-38205 La Laguna, Tenerife, Spain
         \and
             Departamento de Astrof\'isica, Universidad de La Laguna, E-38206 La Laguna, Tenerife, Spain 
        \and
             Instituto de Astronom\'{\i}a, Universidad Nacional Aut\'onoma de M\'exico, Ap. 70-264, 04510 CDMX, M\'exico
        \and
             Dipartimento di Fisica e Astronomia “Augusto Righi”, Alma Mater Studiorum, Università di Bologna, Via Gobetti 93/2, 40129 Bologna, Italy
         \and
            INAF-Osservatorio di Astrofisica e Scienza dello Spazio di Bologna, Via Gobetti 93/3, 40129 Bologna, Italy
        \and
            Max Planck Institute for Astrophysics, Karl-Schwarzschild-Str. 1, D-85748 Garching, Germany
        \and
            Sterrenkundig Observatorium, Ghent University, Krijgslaan 281 - S9, 9000 Gent, Belgium
        \and 
            Dept. F\'isica Te\'orica y del Cosmos, E-18071 Granada, Spain
        \and
            Instituto Universitario Carlos I de F\'isica Te\'orica y Computacional, Universidad de Granada, E-18071 Granada, Spain
        \and 
            Universidad Nacional Aut\'onoma de M\'exico, Instituto de Astronom\'\i a, AP 106, Ensenada 22800, BC, M\'exico
           }

\authorrunning {M\'endez-Delgado et al.}
\titlerunning {Fe/O and Fe/N abundances in Star-Forming Regions}
   \date{\today}

% \abstract{}{}{}{}{} 
% 5 {} token are mandatory
 
  \abstract
   {In stars, metallicity is usually traced using Fe, while in nebulae, O serves as the preferred proxy. Both elements have different nucleosynthetic origins and are not directly comparable. Additionally, in ionized nebulae, Fe is heavily depleted onto dust grains.}
   {We investigate the distribution of Fe gas abundances in a sample of 452 star-forming nebulae with \feiii~$\lambda 4658$ detections and their relationship with O and N abundances. Additionally, we analyze the depletion of Fe onto dust grains in photoionized environments.}
   {We homogeneously determine the chemical abundances with direct determinations of electron temperature ($T_e$), considering the effect of possible internal variations of this parameter. We adopt a sample of 300 Galactic stars  to interpret the nebular findings.}
   {We find a moderate linear correlation ($r=-0.59$) between Fe/O and O/H. In turn, we report a stronger correlation ($r=-0.80$) between Fe/N and N/H. We interpret the tighter correlation as evidence of Fe and N being produced on similar timescales while Fe-dust depletion scales with the Fe availability. The apparently flat distribution between Fe/N and N/H in Milky Way stars supports this interpretation. We find that when 12+log(O/H)<7.6, the nebulae seem to reach a plateau value around $\text{log(Fe/O)} \approx -1.7$. If this trend is confirmed, it would be consistent with a very small amount of Fe-dust in these systems, similar to what is observed in high-z galaxies discovered by the \textit{James Webb Space Telescope} (JWST). We derive a relationship that allows us to approximate the fraction of Fe trapped into dust in ionized nebulae. If the O-dust scales in the same way, its possible contribution in low metallicity nebulae would be negligible. After analyzing the Fe/O abundances in J0811+4730 and J1631+4426, we do not see evidence of the presence of very massive stars with $M_\text{init}>300M_{\odot}$ in these systems.}
   {The close relation observed between the N and Fe abundances has the potential to serve as a link between stellar and nebular chemical studies. This requires an expansion of the number of abundance determinations for these elements in both stars and star-forming nebulae, especially at low metallicities.}


   \maketitle
%
%-------------------------------------------------------------------

\section{Introduction}
\label{sec:intro}

The determination of metallicity, which refers to the abundance of elements heavier than helium, and its distribution are key to understanding the formation and evolution of galaxies in the Universe. However, studies focused in the chemical composition of the ionized nebulae trace metallicity using the oxygen abundance, whereas the stars typically use iron as a proxy for this parameter. Since the nucleosynthetic origin of both elements is different, the comparison between one metallicity proxy and the other is not straightforward, being the central topic of several studies \citep{Matteucci:86,Wheeler:89,Gratton:00,Walcher:09,Nicholls:17,Sanchez:21,Chruslinska:23}. 

O is produced primarily by massive stars and mainly released into the interstellar medium (ISM) through core-collapse supernovae (CCSN) on short time-scales \citep{Woosley:02, Chiappini:03, Kobayashi:20}. Other elements like Ne, S, and Ar can also be produced by massive stars through the alpha process, which would link their relative abundances \citep{Chiappini:03,Carigi:05,Croxall:16, Esteban:20, Rogers:22, Arellano:24}. On the other hand, the Fe-peak elements such as Fe, Ni, Cr, Mn are mostly produced by explosive nucleosynthesis, mainly in type Ia supernovae (SN-Ia) \citep{Chiappini:03} with variable contributions from massive stars \citep{Gratton:00, Palla:21}. Due to these differences, it is expected that the abundance of Fe is not always proportional to that of O, but  their relative abundance depends on the stellar formation and evolution. For instance, it is expected that in young regions of high star formation, the production of Fe would be delayed in comparison to that of O, since the former element requires a time period of at least $\sim 40$ Myrs to form stellar systems that give rise to SN-Ia \citep{Greggio:05, Maoz:12, Chruslinska:23}.

Nitrogen is an interesting element that could serve as a bridge between stellar metallicities, usually traced via Fe, and nebular ones, commonly traced via O. Several works have pointed to two main nucleosynthetic mechanisms as the origin of N \citep{Vila-Costas:93, Henry:00, Israelian:04, Nava:06, Nicholls:17, Romano:19, Grisoni:21}. N can be produced on short time-scales, analogous to O, known as the primary production mechanism. On the other hand, intermediate-mass stars ($3M_{\odot}<M<8M_{\odot}$) are capable of producing N at the expense of C and O through the CNO cycle \citep{Hoek:97, Henry:00, Vincenzo:16,Ventura:22}, known as the secondary production mechanism. Stars more massive than those mentioned earlier may also contribute nitrogen formed through the CNO cycle \citep{Molla:06, Przybilla:10}, although their overall contributions are expected to be smaller \citep{Henry:00}. Wolf-Rayet stars can also contaminate the ISM with nitrogen-rich ejecta through stellar winds \citep{Meynet:2005, Crowther:07}. The release of the main secondary-produced N requires timescales that allow the evolution of intermediate-mass stars ($\sim 40$ Myr \citep{Chruslinska:23}). This condition is also necessary for the formation of stellar systems that give rise to SN-Ia, producers of Fe.

%Observationally, the relationships between Fe, N and O (or other nucleosynthetically linked $\alpha$-elements) have been mostly studied through stellar signatures \citep{Gratton:00, Cayrel:04, Bensby:13} with a smaller number of works dedicated to ionized nebulae \citep{Rodriguez:05, Izotov:06, Peimbert:10, Kojima:21}. 

There are various difficulties in analyzing Fe, N, and O relations in the nebular emission spectra of star-forming regions. In these systems, most of the Fe atoms are usually depleted into dust grains, which are able to survive in photoionized environments \citep{Osterbrock:92, Rodriguez:96,Izotov:06, Roman-Duval:22, Roman-Duval:22a}. \feiii~emission lines arising from the gaseous-phase iron are typically faint, up to an order of magnitude fainter than temperature-sensitive auroral lines (e.g., \oiii~$\lambda4363$, \nii~$\lambda5755$). Other Fe-signatures, such as those from \feii, can arise from starlight through fluorescent processes, further complicating their analysis \citep{Rodriguez:99, Verner:00}. Additionally, the complex atomic structures of the Fe ions make their radiative and collisional models under nebular conditions susceptible to containing significant systematic uncertainties \citep{Rodriguez:05}.

Fortunately, since the pioneering work of \citet{Osterbrock:92}, who determined the first gaseous Fe/H abundance in the Orion Nebula, there has been significant progress both observationally and theoretically. Studies dedicated to the properties of the faint auroral lines and ultra-faint heavy element recombination lines (RLs) have indirectly also detected multiple \feiii~emission lines \citep{Rodriguez:02, Izotov:06, Esteban:09, Peimbert:07, MesaDelgado:09, Berg:15, Kojima:21, MendezDelgado:21a}. On the other hand, recent atomic studies have allowed for the inference of the Fe$^{2+}$ radiative and collisional properties with an accuracy of $\sim 20\%$ \citep{Mendoza:23}, sufficient for reliable estimates of its ionic abundances. This opens up the possibility of accurately studying the gaseous Fe abundance in star-forming regions and its relationship with other chemical properties. Since \feiii~lines are present even in some high-redshift galaxies, recently unveiled with the \textit{James Webb Space Telescope} (JWST) \citep{ArellanoCordova:22, Welch:24}, a better understanding of this element provides additional pieces of information about the composition and chemical evolution of the Universe.

The depletion of Fe into dust grains in photoionized environments is an active research area, with important implications for other elements such as O \citep{Jenkins:09,Jones:17,Hensley:23}. There is no consensus on the precise chemical composition of dust in photoionized environments, but it is expected to consist mainly of the leftovers from the cool clouds where star formation initially started. In neutral clouds, it is believed that Fe exists as nano-inclusions in large silicate dust grains and other types of free-flying Fe compounds \citep{Jones:17,Hensley:23}. If most of the Fe-rich dust grains present in photoionized environments are in the form of silicates, they could then be linked to the depletion of O. In contrast, the survival of mostly free-flying Fe compounds would decouple the depletion of Fe from that of O. Therefore, determining the Fe dust fraction and its environmental dependencies may be key to quantifying the O depletion and understanding the properties of interstellar dust.

Additionally, recent results based on the high gaseous abundances of Fe/O in low-metallicity galaxies by \citet{Kojima:21} have suggested the existence of very massive stars with $M_\text{init}>300M_{\odot}$ in these systems. The presence of such stars would imply the need for inclusion in photoionization models \citep{Goswami:21,Goswami:22,Watanabe:24}, which in turn could help us to understand the existence of intense \heii~lines in low-metallicity photoionized systems \citep{LopezSanchez:09,Schaerer:19}. 

Given the importance of the gas phase Fe abundances and its relation with other elements as O and N in several physical phenomena, we examine in this study the largest sample of star-forming regions, both Galactic and extragalactic, from the literature with precise determinations of Fe, O, and N. We further adopt stellar chemical abundances from the literature to compare and discuss in the context of the results from ionized nebulae. Following the methodology of the DEep Spectra of Ionized REgions Database (DESIRED) project \citep{MendezDelgado:23b}, we have uniformly analyzed 452 optical spectra from Galactic and extragalactic star-forming regions, including the Sunburst Arc, a high-redshift system ($z=2.37$), recently observed with the JWST \citep{Welch:24}. We have directly determined their ionic abundances of Fe$^{2+}$/H$^{+}$, N$^{+}$/H$^{+}$, O$^{+}$/H$^{+}$, O$^{2+}$/H$^{+}$, through precise calculations of their electron density and temperature ($n_e, T_e$), while also considering the effects of temperature inhomogeneities \citep{Peimbert:67,Peimbert:69,Bergerud:20,Cameron:23,MendezDelgado:23a}.

%We present the relationship between gas-phase Fe/O abundance and metallicity (traced by O) and a stronger correlation between gas-phase Fe/N abundances and N/H. We derive empirical relationships to estimate the fraction of Fe and O trapped into dust and discuss their implications.

In Sec.~\ref{sec:obs}, we describe our observational sample of both stars and nebulae. In Sec.~\ref{sec:physical_chemical}, we outline the methodology used to calculate the physical conditions and chemical abundances of the analyzed nebular sample. Additionally, Sec.~\ref{sec:ICF_Fe_errors} is dedicated to detailing the issues involved in calculating the total gaseous abundance of Fe, owing to discrepancies between photoionization models and direct determinations of Fe$^{3+}$ abundance. In Sec.~\ref{sec:results}, we present the nebular and stellar Fe/O and Fe/N distributions from our sample. These results and their implications for various astrophysical topics are discussed in Sec.~\ref{sec:discussion}. Finally, in Sec.~\ref{sec:concl}, we summarize our findings.

%--------------------------------------------------------------------
\section{Observational sample}
\label{sec:obs}
%-------------------------------------- Two column figure (place early!)

As part of the DESIRED project \citep{MendezDelgado:23b}, we compile all reported emission line intensities, corrected by reddening, from Table~\ref{table:IDs}. This constitutes a subset of what an extended version of DESIRED encompasses ($\sim2000$ optical spectra with direct determinations of $T_e$). We standardize the formats, providing individual notes for each line in case of observational defects or blends. This enables a uniform and consistent analysis of the entire database in a straightforward fashion. Analysis codes and the standardized spectra will be presented in a forthcoming paper. DESIRED aims to be a collaborative project that facilitates the analysis of deep spectra from photoionized regions, including \hii~regions, planetary nebulae and Herbig-Haro objects beyond partial compilations of a limited number of emission lines. 

The extension of the original DESIRED sample \citep{MendezDelgado:23b} brings additional complications due to the different observational conditions present in the literature. For example, at low spectral resolution, lines typically used in density diagnostics such as \ariv~$\lambda 4711$ can be blended with \hei~$\lambda 4713$, or other temperature-sensitive lines like \oiii~$\lambda 4363$ can be mixed with \feii~$\lambda 4360$. In all these cases, we have placed special emphasis and added cautionary notes when compiling this information in the DESIRED format. These notes are considered to discard lines with observational problems from the analysis of physical conditions, analogous to the approach established by \citet{MendezDelgado:23b}. In addition to handling notes, we consider additional tests to ensure the quality of the data. For example, we verify that \oiii~$\lambda 5007/\lambda 4959$, \siii~$\lambda 9531/\lambda 9069$, and \nii~$\lambda 6584/\lambda6548$ are consistent with theoretical values within 20\%; otherwise, they are discarded. This is described in higher detail in Sec.~\ref{sec:physical_chemical}.

\begin{figure}[h]
\centering    
\includegraphics[width=\hsize ]{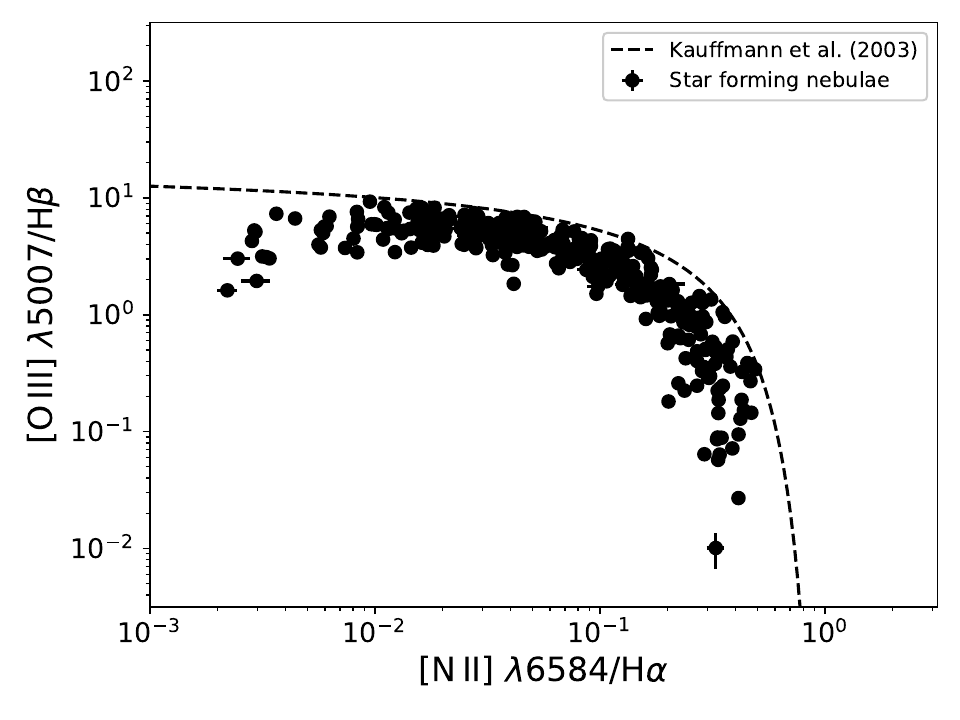}
\caption{BPT diagram of the selected nebular spectra. The dashed line represents the empirical relation by \citet{Kauffmann:03} that distinguishes between star-forming regions and active galactic nuclei (AGNs).} 
\label{fig:BPT_diagram}
\end{figure}

The selected data for this article present reliable detections (with errors smaller than 40\%) of the \feiii~$\lambda 4658$ line, which is the most suitable line for determining the Fe$^{2+}$/H$^{+}$ abundances due to its relatively high intensity \citep{Rodriguez:02, MesaDelgado:09, MendezDelgado:21a} and because it originates from some of the best-known atomic transitions \citep{Mendoza:23}. Additionally, a sub-sample of the data contains detections of \feiii~$\lambda 4702$, another relatively bright and well-known line. More importantly, these observations contain at least one detection of the following auroral/nebular intensity ratios: \oiii~$\lambda4363/\lambda 5007$, \nii~$\lambda5755/\lambda6584$, and \siii~$\lambda 6312/\lambda9069$. These ratios are highly sensitive to $T_e$ and are insensitive to density up to values of $n_e\sim 10^4 \text{ cm}^{-3}$ \citep{FroeseFischer:04,Tayal:11}. A good determination of $T_e$ is critical for obtaining reliable chemical abundances in optical spectra, given the exponential dependence of line emissivity of the collisionally excited lines (CELs) on this parameter \citep{Osterbrock:06}. In Fig.~\ref{fig:BPT_diagram} we show the position of the sample of \hii~regions used in the present study in the Baldwin-Phillips-Terlevich (BPT) diagram \citep{Baldwin:81}, showing that we have not included regions with hard ionizing sources, which are usually to the right of the \citet{Kauffmann:03} line. The observational sample covers a metallicity range from 12+log(O/H)$\approx6.9$ to 12+log(O/H)$\approx8.9$ \citep[$t^2=0$,][]{Peimbert:67}.

%We also take values for B-type stars from the Small Magellanic Cloud (SMC) \citep{Trundle:05, Hunter:07}, the Large Magellanic Cloud (LMC) \citep{Hunter:07}, and IC1613 \citep{Bresolin:07b}

Additionally, we adopt Galactic stellar abundances available for Fe, O, and N from B-type stars \citep{Nieva:12, Wessmayer:2022} and other dwarf stars, including  metal-poor ones \citep{Carretta:00, Israelian:04, Cayrel:04, Ecuvillon:04, Spite:06, Bensby:13, Magrini:18, Amarsi:19}. We exclude stars that have undergone mixing with deep layers of the star, thereby affecting the observed CNO abundances from \citet{Carretta:00} and \citet{Spite:06}. We add a 0.40 dex correction to the Fe abundances in the stars from \citet{Spite:06} for 3D/NLTE effects \citep{Spite:06, Kobayashi:20}.

The selected B-type stars represent the present-day chemical composition of the ISM, and their elemental abundances could ideally be compared to those of \hii~regions. However, in practice, this is not the case. Detailed studies of the O/H abundances from ionizing stars in Galactic \hii~regions \citep{SimonDiaz:06, SimonDiaz:11, GarciaRojas:14} show systematically higher abundances of $\sim 0.2 \text{ dex}$ in comparison to the nebular determinations when using CELs and the so-called ``direct method'' \citep{Dinerstein:90, Peimbert:17}. If instead of nebular CELs, ultra-faint heavy element RLs are used, the discrepancies practically disappear, which is interpreted as evidence in favor of an inhomogeneous nebular temperature structure \citep{Peimbert:67, Carigi:05, GarciaRojas:07b, MendezDelgado:22, Esteban:22}. Other giant B-type stars may exhibit large chemical variations in CNO abundances due to strong mixing processes, complicating their direct comparison with \hii~regions element by element \citep{Trundle:07, Hunter:07,Przybilla:10,Garcia:14}. This will be further discussed in Sec.~\ref{subsec:very_massive}.The metal-poor Galactic stars do not reflect the current composition of the ISM, but rather part of its chemical past if they are not polluted by their own evolution. This is the case with our selected sample of dwarf stars. Their inclusion allows us to understand the evolution of Fe/O and Fe/N abundances with respect to O/H and N/H.

Overall, the number of low-metallicity stars with simultaneous determinations of N and Fe in the literature is relatively limited. A primary challenge often encountered is the detection of N abundance indicators, which remain faint even at solar metallicities \citep{Asplund:05,Amarsi:20, Amarsi:21}. Additionally, the determination of Fe can be problematic, particularly in the case of hotter massive stars. For stars with $T_\text{eff} \gtrsim 24\,$kK, corresponding approximately to the spectral type B0.5, there are no optical Fe lines \citep[e.g.,][]{Thompson:08} and instead the iron forest at UV wavelengths has to be analyzed, which is challenging. Even in the B-star regime, the Fe lines are generally weak and high-quality (high S/N, high resolution) spectra are required. Hence, many studies examining abundances in massive stars resort to determine CNO abundances, but adopt a solar or solar-scaled Fe abundance by default \citep[e.g.,][]{Martins:15, Braganca:19}.

%--------------------------------------------------------------------
\section{Nebular physical conditions and chemical abundances}
\label{sec:physical_chemical}
%-------------------------------------- Two column figure (place early!)

%--------------------------------------------------- One column table

\subsection{Electron density and temperature}
\label{subsec:ne_Te}

To determine the physical conditions of the gas, we use \textit{PyNeb 1.1.18} \citep{Luridiana:15} with the atomic dataset presented in Table~\ref{table:atomic_data} and the \hi~effective recombination coefficients from \citet{Storey:95}. We first employ the \textit{getCrossTemDen} routine with line intensity ratios sensitive to $n_e$ and $T_e$. We test density-sensitive diagnostics such as \sii~$\lambda 6731/\lambda6716$, \oii~$\lambda 3726/\lambda3729$, \cliii~$\lambda 5538/\lambda5518$, \feiii~$\lambda 4658/\lambda4702$, and \ariv $\lambda 4740/\lambda4711$ with temperature-sensitive diagnostics including \nii~$\lambda 5755/\lambda 6584$, \oiii~$\lambda 4363/\lambda 5007$, \ariii~$\lambda 5192/\lambda 7135$, and \siii~$\lambda 6312/\lambda 9069$.

Firstly, each density-sensitive diagnostic is cross-correlated with all available temperature diagnostics using a Monte Carlo experiment of 100 points to propagate uncertainties in line intensities. For each convergence, a value of $n_e$ and $T_e$ and their associated uncertainties are obtained. The average $n_e$, weighted by the inverse square of the error of the different convergences, is adopted as the representative value of the diagnostic (e.g., $n_e$(\oii~$\lambda 3726/\lambda3729$)). This approach enables us to consider the small temperature dependence of density diagnostics under typical nebular conditions.

Once the density for each diagnostic is established, we follow the criteria suggested in Sec.~5 by \citet{MendezDelgado:23b} to estimate a global average density. If $n_e$(\sii~$\lambda 6731/\lambda6716) < 100\text{ cm}^{-3}$, we adopt $n_e=100 \pm 100\text{ cm}^{-3}$. If $100 \text{ cm}^{-3}\leq n_e$(\sii~$\lambda 6731/\lambda6716)$ $< 1000\text{ cm}^{-3}$, we adopt the average between $n_e$(\sii~$\lambda 6731/\lambda6716$) and $n_e$(\oii~$\lambda 3726/\lambda3729$). If $n_e$(\sii~$\lambda 6731/\lambda6716)\geq 1000\text{ cm}^{-3}$, we adopt the averages of $n_e$(\sii~$\lambda 6731/\lambda6716$), $n_e$(\oii~$\lambda 3726/\lambda3729$), $n_e$(\cliii~$\lambda 5538/\lambda5518$), $n_e$(\feiii~$\lambda 4658/\lambda4702$), and $n_e$(\ariv~$\lambda 4740/\lambda4711$). In cases where it is not possible to determine the density, such as in the spectra reported by \citet{Izotov:06}, we adopt $n_e=100 \pm 100\text{ cm}^{-3}$.

Although there is temperature stratification \citep[i.e., a representative temperature for each ionization volume,][]{Stasinska:78, Osterbrock:06, Peimbert:17, Berg:21}, the possible existence of density stratification is not obvious among the different density diagnostics of \hii~regions \citep{MendezDelgado:23b}. These diagnostics have non-uniform sensitivities with $n_e$, and are susceptible to systematic biases \citep{peimbert:71, Rubin:89, Tsamis:03, MendezDelgado:22}. The impact of these density biases is small in optical spectra if a good determination of $T_e$ is adopted \citep[up to 0.1 dex if auroral lines are used to determine ionic abundances;][]{MendezDelgado:23a, RickardsVaught:24} but critical in chemical abundance studies with fine structure infrared CELs \citep{Rubin:89, Tsamis:03, Stasinska:13, MendezDelgado:24}. Therefore, as a good approximation, adopting the global average density value, we estimate the temperatures $T_e$(\nii~$\lambda 5755/\lambda 6584$), $T_e$(\oiii~$\lambda 4363/\lambda 5007$), $T_e$(\ariii~$\lambda 5192/\lambda 7135$), and $T_e$(\siii~$\lambda 6312/\lambda 9069$) using the \textit{getTemDen} routine from PyNeb. 

Being a critical parameter, we exercise strict control over the determination of $T_e$. In addition to excluding lines with errors larger than 40\%, we compare the observed line intensity ratios of nebular transitions with their theoretical predictions. Lines arising from the same upper atomic level (e.g., \oiii~$\lambda \lambda 5007, 4959$, \siii~$\lambda \lambda 9531, 9069$, \nii~$\lambda \lambda 6584, 6548$) have fixed relative intensities given by the Einstein radiative coefficients, regardless of the physical conditions of the gas \citep{Storey:00}. If the observed line intensity ratios differ by more than 20\% from the theoretical predictions, their use for determining $T_e$ is discarded. In the particular case of \siii~$\lambda \lambda 9531, 9069$, when $\lambda 9531/\lambda 9069 < 2.47$, it is assumed that \siii~$\lambda 9531$ is affected by telluric absorption bands \citep{Noll:12}, while \siii~$\lambda 9069$ is not. This is the most common case in the literature with particular exceptions such as the Orion Nebula \citep{Baldwin:96}. Unfortunately, in a small number of cases, it is not possible to perform this test, as only one of the nebular lines is reported. In these cases, we consider the $T_e$ results valid.  In any case, for this specific work, we only adopt T(\siii) to determine ionic abundances in the absence of T(\nii) and T(\oiii). 

Another important case is that of $T_e$(\oiii), where \oiii~$\lambda 4363$ can be blended with \feii~$\lambda 4360$ \citep{Curti:17} when the spectral resolution is intermediate or low. In instances where we have detected this blend, we have discarded the use of \oiii~$\lambda 4363$, ensuring that we do not introduce spurious overestimations of $T_e$(\oiii). The impact of this blend is greater in regions of high metallicity and low ionization degree \citep{ArellanoCordova:2020b}. Most of the star-forming regions with these characteristics analyzed in this work come from the CHAOS sample \citep{Berg:15}, which has taken special care in separating \oiii~$\lambda 4363$ and \feii~$\lambda 4360$ \citep{Berg:20,Rogers:21,Rogers:22}. Therefore, this potential observational issue will not systematically affect the analysis of the sample presented here. We show our derived physical conditions in Tables \ref{table:densities} and \ref{table:temperatures}.

%Some examples of discarded auroral to nebular intensity ratios include NGC5408-2 \citep{Guseva:11} where \oiii~$\lambda 5007/4959=4.16$ (theoretical value $3.00$), +80d0-148d2 \citep{Berg:20} where \siii~$\lambda 9531/9069=3.64$ (theoretical value $2.47$), or NGC346 \citep{Valerdi:19} \nii~$\lambda 6584/6548=1.96$ (theoretical value $3.05$).

\subsection{Ionic abundances}
\label{subsec:ion_ab}

To determine the ionic abundances of Fe$^{2+}$, N$^{+}$, O$^{+}$, and O$^{2+}$, we use the \feiii~$\lambda \lambda 4658, 4702$, \nii~$\lambda \lambda 6548, 6584$, \oii~$\lambda \lambda 3727, 3729$, and \oiii~$\lambda \lambda 4959, 5007$ lines. In cases where both lines of each pair are available, we use the sum of their relative intensities to H$\beta$. Otherwise, we use the available line. In cases where \oii~$\lambda \lambda 3727, 3729$ are not available due to spectral coverage limitations or observational defects in the blue arm, we use the auroral lines \oii~$\lambda \lambda 7319, 7320, 7330, 7331$ to estimate the abundance of O$^{+}$. Considering similarities in ionization potential, we adopt a common temperature, representative of low-ionization ions, to estimate the abundances of Fe$^{2+}$, N$^{+}$, and O$^{+}$. In the same manner, to determine the abundance of O$^{2+}$, we adopt a temperature representative of high-ionization.

Ionic abundances are estimated in two ways: considering a nebular homogeneous temperature structure ($t^2=0$), known as ``the direct method'', and accounting for the effect of an inhomogeneous temperature structure ($t^2>0$) \citep{Peimbert:67}. Following the empirical results of \citet{MendezDelgado:23a}, we performed corrections for temperature inhomogeneities only for highly ionized ions. For $t^2=0$, we adopt $T_e$(\nii~$\lambda 5755/\lambda 6584$) and the global $n_e$ in the \textit{getIonAbundance} routine of PyNeb, propagating uncertainties through 100-point Monte Carlo experiments. When $T_e$(\nii~$\lambda 5755/\lambda 6584$) is unavailable, we employ the temperature relations of \citet{Garnett:92} to estimate $T_e$(\nii) from $T_e$(\oiii~$\lambda 4363/\lambda 5007$) or $T_e$(\siii~$\lambda 6312/\lambda 9069$), when the first diagnostic is also absent. Similarly, the abundance of O$^{2+}$ is determined using $T_e$(\oiii~$\lambda 4363/\lambda 5007$) and the derived global average density. In cases where $T_e$(\oiii~$\lambda 4363/\lambda 5007$) is not available, we use the temperature relations of \citet{Garnett:92} along with measurements of $T_e$(\nii~$\lambda 5755/\lambda 6584$) and/or $T_e$(\siii~$\lambda 6312/\lambda 9069$). For $t^2>0$, the treatment of low-ionization ions is identical to that described above. In the case of O$^{2+}$, we adopt $T_0$(O$^{2+}$), estimated from Equation (4) of \citet{MendezDelgado:23a}. To roughly estimate the effect of $t^2$ when $T_e$(\nii~$\lambda 5755/\lambda 6584$) is not available, we determine $T_0$(O$^{2+}$) by combining Equation (4) of \citet{MendezDelgado:23a} with Equation (2) of \citet{Garnett:92}. 

In the present study, we do not add additional errors when using a temperature relationship to connect one $T_e$ diagnostic with another (e.g., if we have $T_e$(\nii) with a 15\% error, and we use a relationship to infer $T_e$(\oiii), we assume that $T_e$(\oiii) also has a 15\% error). This is a lower limit to the uncertainties of the inferred temperature, similar to the approach of \citet{Skillman:03}. However, we have been particularly careful about the impact of potential systematic errors introduced by the temperature relationships, with special emphasis on what happens in the low-metallicity regime. Under such conditions, the temperature relationships have not been sufficiently tested with empirical determinations, having limited statistics among simultaneous determinations of $T_e$(\nii), $T_e$(\oiii), and $T_0$(O$^{2+}$) when $T_e$>13,000K \citep{Esteban:09, Berg:20,ArellanoCordova:20,MendezDelgado:23a,MendezDelgado:23b}. This is analyzed in greater detail in Sec.~\ref{sec:appendix_a0}. In general, the potential impact of severe systematic errors in the temperature relationships could be up to$\sim$0.05 dex in the Fe/O and Fe/N distributions. The adopted temperatures for determining the ionic abundaces are shown in Table~\ref{table:temperatures_adopted}. Ionic abundances are shown in Table~\ref{table:ionic_abundances}.

\subsection{Total abundances}
\label{subsec:tot_ab}

To estimate the total abundance of O/H, we directly sum the contributions of the ionic abundances O$^{+}$/H$^{+}$ and O$^{2+}$/H$^{+}$. The possible contribution of O$^{3+}$/H$^{+}$ was not considered, as it is expected to be very small \citep{Amayo:21}. In extremely metal-poor \hii\ regions (12+log(O/H) $<$ 7.5), ionization conditions may be harder than expected in photoionization models, leading to relatively strong emissions of \heii. Under these conditions, an Ionization Correction Factor (ICF) can be implemented, as is typically done in the study of planetary nebulae \citep[PNe,][]{Torres-Peimbert:77}. However, observational studies on the contribution of O$^{3+}$/H$^{+}$ to the total O/H abundance in metal-poor regions often find values up to the order of 5\% \citep{Izotov:99a, DominguezGuzman:22}. This was also found by \citet{Berg:21} in two Extreme Emission Line Galaxies with direct O$^{3+}$/H$^{+}$ abundances measured from the UV \oivs~lines. This fraction is small for the purposes of this study and does not affect our conclusions. However, other studies such as those dedicated to estimating the He/O fractions must consider it, as they require a precision better than 1\% \citep{Peimbert:74, Pagel:92, Izotov:98, Peimbert:02, Aver:15, Valerdi:19, MendezDelgado:20, Hsyu:20, Kurichin:21,Matsumoto:22}.

%\citep{DelgadoInglada:16, DominguezGuzman:22, Esteban:04, Esteban:09, Esteban:13, Esteban:14, Esteban:17, Esteban:20, GarciaRojas:04, GarciaRojas:05, GarciaRojas:06, GarciaRojas:07, Guseva:11, LopezSanchez:07, MesaDelgado:09, MendezDelgado:21a, MendezDelgado:21b, MendezDelgado:22b, Peimbert:03, Peimbert:12, PenaGuerrero:12, Toribio:16, Valerdi:19}.
\begin{figure}[h]
\centering    
\includegraphics[width=\hsize ]{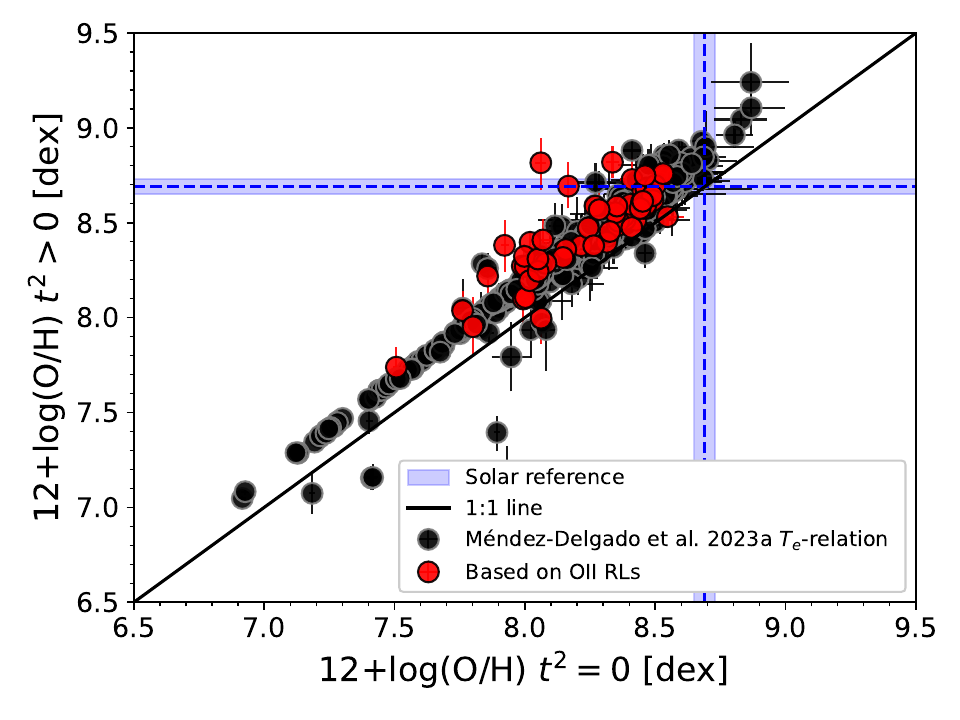}
\caption{Comparison of O/H abundances in the sample of star-forming regions considering a homogeneous temperature structure ($t^2=0$) and considering temperature variations ($t^2>0$) following the empirical relations derived by \citet{MendezDelgado:23a}. Red dots indicate the O/H abundances determined from the ultra-faint \oiirls-RLs, which are insensitive to temperature. The solar O/H abundance from \citet{Asplund:21} is shown as a reference.} 
\label{fig:t2_vs_direct}
\end{figure}

The choice of $t^2=0$ or $t^2>0$ in determining the total O/H abundances of our observational sample of star-forming regions has an impact of 0.1-0.2 dex, as shown in Fig.~\ref{fig:t2_vs_direct}. All determinations shown in black dots are based on CELs, following the procedure described in Sec.~\ref{subsec:ion_ab}. For comparison, the figure displays abundances estimated with the temperature-insensitive \oiirls~RLs (red dots), in objects with the deepest spectra. Additionally, the solar reference by \citet{Asplund:21} is included. The choice of $t^2=0$ or $t^2>0$  results in differences of around 0.1-0.2 dex in the distributions of Fe/O and Fe/N, as discussed in Sec.~\ref{sec:ICF_Fe_errors}. These differences may be relevant for the absolute values of Fe/O and Fe/N. On the other hand, the shape of the Fe/O and Fe/N distributions seems to remain unchanged, as it is shown in Sec.~\ref{sec:appendix_a}.

From the entire sample, 7 regions (Hubble~V \citep{Peimbert:05}, NGC~3603 \citep{GarciaRojas:06}, POX36, SBS0335-052E , J1205+4551 \citep{Izotov:04,Izotov:09, Izotov:17b, Izotov:21b}, Tol1214-277 \citep{Guseva:11} and NGC\,5449-2 \citep{Croxall:16}) showed substantially higher O abundances in the case of $t^2=0$ compared to the case $t^2>0$ using the temperature relation proposed by \citet{MendezDelgado:23a}. This suggests discrepancies between the values of $T_e$(\nii) and $T_e$(\oiii) beyond temperature variations. The most obvious case is NGC\,5449-2 \citep{Croxall:16}, where \nii~$\lambda 5755/\lambda 6584$ suggest a $T_e$(\nii)>40,000K (see Table~\ref{table:temperatures}). This could be a non-reported observational error (\citet{Croxall:16} did not  report $T_e$(\nii) for this region, but they do not mention any observational problem that would necessitate discarding \nii~$\lambda 5755/\lambda 6584$). In the case of NGC~3603, it may be an aperture effect by not covering the entire Galactic nebula. Other regions like J1205+4551 could have overestimations in $T_e$(\nii) due to some phenomenon related to the presence of WR stars enriching the ISM with N \citep{Izotov:21b}. In these cases, we adopt only the $t^2=0$ values for our analysis, exerting no influence on our conclusions.

It is interesting to note that if $t^2=0$ is considered in general, \hii~regions with abundances clearly higher than solar are rather rare, being only P203 and +30.8+139.0 from the M51 galaxy \citep{Croxall:15}, -35d7+119d6 from NGC\,628 \citep{Berg:15} and +75.7+89.1 and -35.7+119.6 from NGC\,3184 \citep{Berg:20}, respectively. This, of course, could be a selection bias of our sample, dedicated to the detection of \feiii~emission lines. However, in the literature, there are only $\sim 20$ additional examples of regions with $T_e$-based metallicity estimations higher than solar (with $t^2=0$) \citep{Castellanos:02,Rosolowsky:08,Bresolin:05,Bresolin:07a,Croxall:15, Croxall:16,Lin:17,Patterson:12,Berg:15, ArellanoCordova:21, Rogers:21,Rogers:22}. In our Galaxy, only Sh~2-48 and Sh~2-53 \citep{ArellanoCordova:21} would reach metallicities close to solar, but they are approximately 4 kpc closer to the Galactic center than the Sun. In contrast, massive O and B-type stars in the solar neighborhood ($d<3$ kpc) typically exhibit solar or supersolar metallicities \citep{SimonDiaz:06, Nieva:12, Martins:15, Wessmayer:2022}.

%This, of course, could be a selection bias of our sample, dedicated to the detection of \feiii~emission lines. However, in the literature, there are only a few additional examples of regions with $T_e$-based metallicity estimations higher than solar. To our knowledge, there are only $\sim 20$ cases, from the following galaxies: M101: H493 \citep{Bresolin:07a}, H760, H864 \citep{Croxall:16}; M33: B0015c \citep{Rosolowsky:08}, BCLMP218B, GDK99-118, NGC604C \citep{Lin:17}, +02+111 \citep{Rogers:22}; M51: -33d2+58d0, -78d9+107d4, -97d0-78d4, +91d0+69d0 \citep{Croxall:15}; M81: disc1 \citep{Patterson:12}; M83: 11 \citep{Bresolin:05}; NGC1232: 5 \citep{Bresolin:05}, CDT1 \citep{Castellanos:02}; NGC2403: VS35 \citep{Rogers:21}; NGC2997: 5\citep{Bresolin:05}, NGC628: +61d2+113d5, +81d6-32d3 \citep{Berg:15}. Esto contrasta 

To determine the abundances of N/H and Fe/H, the use of an ICF is required to estimate the contributions of unobserved ionic states. In the case of N, it is necessary to estimate the contribution of N$^{2+}$ in the ionized gas that does not emit lines in the optical spectrum. To this purpose, we adopt the scheme proposed by \citet{Amayo:21}, based on photoionization models by \citet{ValeAsari:16} constructed using CLOUDY \citep{Ferland:17}. The ICF of \citet{Amayo:21} utilizes the similarity between the ionization potentials of N$^{+}$ and O$^{+}$ as proposed by \citet{Peimbert:69}, while also considering departures dependent on the ionization degree of the gas.

In the case of Fe, it is necessary to consider both the contribution of Fe$^{+}$ and Fe$^{3+}$. Although there are lines such as \feii~$\lambda 8617$ or \feiv~$\lambda 6740$ optimal for the direct estimation of ionic abundances, these are generally very weak or found in complicated spectral regions affected by telluric absorptions. Therefore, initially one would think of using a classical ICF as those adopted in the case of other elements such as N, S, Ar, Ne \citep{Amayo:21}. Considering the ionization potentials of Fe$^{2+}$, O$^{+}$, and N$^{+}$, the obvious choice would be to construct an ICF scheme considering the predictions of photoionization models on the relative abundances of these ions. However, Fe is a particularly complex element in star-forming nebulae because it is heavily depleted in dust grains. The relationships between the ionic fractions of Fe may be inconsistent with photoionization models if the fraction of Fe depleted in dust grains is different in the volumes where Fe$^{+}$, Fe$^{2+}$, and Fe$^{3+}$ ions are located. We generally adopt the ICF proposed in Eq.~2 of \citet{Rodriguez:05}. However, it is important to note that the Fe-ICF is a potentially important source of errors as is discussed in Sec.~\ref{sec:ICF_Fe_errors}. ICF(N) and ICF(Fe) numerical values are presented in Table~\ref{table:ICF_numerical} while the total abundances are shown in Table~\ref{table:total_abundances}. 

\section{The problematic calculation of the total gaseous Fe abundances in ionized nebulae}
\label{sec:ICF_Fe_errors}

\begin{figure}[h]
\centering    
\includegraphics[width=\hsize ]{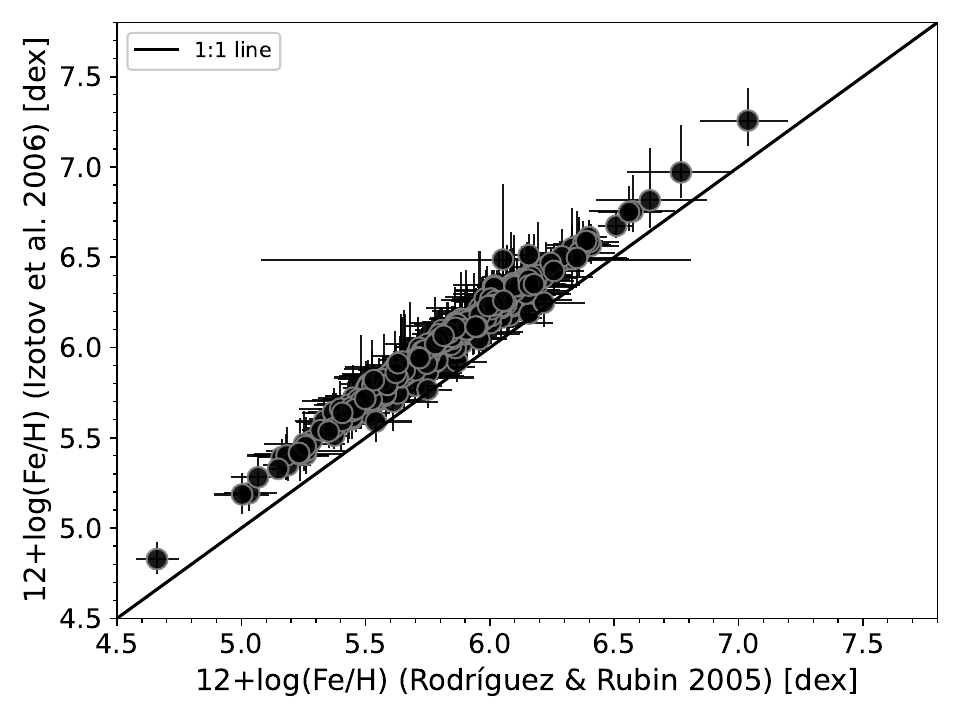}
\caption{Comparison between the Fe-ICF models of \citet{Izotov:06} and \citet{Rodriguez:05} in our adopted sample of star-forming nebulae. Both schemes were derived from photoionization models and are based on the determination of the ionic abundance of Fe$^{2+}$. } 
\label{fig:Comparing_ICFs}
\end{figure}

It is well known that different ICF schemes in the literature can yield significantly different results for various elements \citep[e.g., see the discussions by][]{ArellanoCordova:20, Amayo:21, Arellano:24}. In the case of Fe, there is a systematic discrepancy between the predictions of Eq.~24 from \citet{Izotov:06} and Eq.~2 from \citet{Rodriguez:05}. In Fig.~\ref{fig:Comparing_ICFs}, we show the comparison between both schemes in our sample of star-forming nebulae. The differences between these two ICFs for Fe can reach a systematic offset of up to $\sim 0.2$ dex \citep[see also][] {Kojima:21}. To determine whether the ICF model of \citet{Izotov:06} overestimates the Fe abundances or the model of \citet{Rodriguez:05} underestimates them, or even the possibility that both models provide incorrect predictions, it is necessary to test both models in regions where direct estimates of all ionization states of Fe expected in the ionized gas are available.

\begin{table*}
    \caption{Ionic and total Fe abundances compared to the predictions of the Fe-ICFs from \citet{Rodriguez:05} and \citet{Izotov:06}, derived from photoionization models. These values were calculated both considering an homogeneous nebular temperature ($t^2=0$) and considering the effect of $t^2>0$. The abundance values are presented in units of 12+log(X/H). The reference number is also shown for consistency with the values in Table \ref{table:IDs}.}
    \label{table:total_Fe_direct}
    \begin{tabular}{lllllllllll}
        \hline
        \noalign{\smallskip}
        Ref. & Region  & Fe$^{+}$/H$^{+}$ & Fe$^{2+}$/H$^{+}$ & Fe$^{3+}$/H$^{+}$ & Fe$^{4+}$/H$^{+}$ & Ionic Sum & ICF & ICF \\
         & & & & & & &  R\&R05$^{g}$ &  I06$^{h}$\\
        \noalign{\smallskip}
        \hline
        \multicolumn{9}{c}{$t^2=0$}\\

D83 & HS1851+6933$^{a}$ &  -  & $5.20 ^{+0.06} _{-0.05}$ & $5.59 \pm 0.04$ & $4.49 ^{+0.05} _{-0.04}$ &$5.76 \pm 0.03$ &$5.88 \pm 0.06$ &$6.12 ^{+0.06} _{-0.05}$ \\

D451 & W1702+18$^{a}$ &   -  & $4.95 ^{+0.04} _{-0.03}$ & $5.28 ^{+0.06} _{-0.05}$ & $4.53 \pm 0.05$ &$5.50 \pm 0.04$ &$5.81 \pm 0.03$ &$6.07 \pm 0.03$ \\

D224 & Orion Nebula$^{b}$& $4.69 ^{+0.04} _{-0.05}$ & $5.55 \pm 0.03$ & $5.71 \pm 0.07$ &  -  &$5.96 \pm 0.04$ &$6.11 \pm 0.03$ &$6.32 \pm 0.03$ \\

D235 & NGC2579$^{c}$  &  -  & $5.40 \pm 0.03$ & $5.24 ^{+0.16} _{-0.14}$ &  -  &$5.63 ^{+0.06} _{-0.07}$ &$5.88 \pm 0.03$ &$6.11 \pm 0.03$ \\

D236 & NGC3576$^{d}$  & $4.59 \pm 0.06$ & $5.53 ^{+0.04} _{-0.03}$ & $5.72 ^{+0.09} _{-0.08}$ &  -  &$5.95 \pm 0.05$ &$5.95 \pm 0.03$ &$6.18 ^{+0.04} _{-0.03}$ \\

D283 & Mrk71$^{e}$ &  -  & $4.32 ^{+0.18} _{-0.10}$ & $4.97 ^{+0.08} _{-0.07}$ &  -  &$5.06 \pm 0.06$ &$5.45 \pm 0.14$ &$5.72 ^{+0.18} _{-0.10}$ \\

D403 & SMC-N88A$^{f}$ &   -  & $4.99 ^{+0.04} _{-0.03}$ & $5.47 \pm 0.04$ &  -  &$5.59 \pm 0.03$ &$6.00 ^{+0.08} _{-0.06}$ &$6.27 ^{+0.04} _{-0.03}$ \\

\multicolumn{9}{c}{$t^2>0$}\\
        
D83 & HS1851+6933$^{a}$  & -  & $5.20 ^{+0.07} _{-0.05}$ & $6.48 ^{+0.13} _{-0.09}$ & $5.05 ^{+0.09} _{-0.06}$ &$6.52 ^{+0.11} _{-0.10}$ &$6.29 \pm 0.06$ &$6.58 ^{+0.06} _{-0.05}$ \\

D451 & W1702+18$^{a}$ &   -  & $4.94 \pm 0.03$ & $5.63 ^{+0.07} _{-0.05}$ & $4.74 ^{+0.05} _{-0.04}$ &$5.76 ^{+0.05} _{-0.04}$ &$5.98 \pm 0.03$ &$6.25 \pm 0.03$ \\

D224 & Orion Nebula$^{b}$& $4.69 ^{+0.05} _{-0.04}$ & $5.55 \pm 0.03$ & $5.75 ^{+0.10} _{-0.09}$ &  -  &$5.99 \pm 0.05$ &$6.12 \pm 0.03$ &$6.34 \pm 0.03$ \\

D235 & NGC2579$^{c}$  &  -  & $5.40 \pm 0.03$ & $5.84 \pm 0.14$ &  -  &$5.97 \pm 0.10$ &$6.10 \pm 0.03$ &$6.31 \pm 0.03$ \\

D236 & NGC3576$^{d}$  & $4.59 \pm 0.05$ & $5.53 ^{+0.04} _{-0.03}$ & $6.42 ^{+0.13} _{-0.09}$ &  -  &$6.48 \pm 0.09$ &$6.19 \pm 0.04$ &$6.40 \pm 0.03$ \\

D283 & Mrk71$^{e}$ &  -  & $4.33 ^{+0.17} _{-0.11}$ & $5.49 ^{+0.47} _{-0.17}$ &  -  &$5.51 ^{+0.31} _{-0.29}$ &$5.70 \pm 0.14$ &$6.03 ^{+0.19} _{-0.11}$ \\

D403 & SMC-N88A$^{f}$ &  -  & $4.99 ^{+0.04} _{-0.03}$ & $5.78 ^{+0.09} _{-0.06}$ &  -  &$5.84 ^{+0.07} _{-0.06}$ &$6.16 \pm 0.04$ &$6.44 ^{+0.04} _{-0.03}$ \\
\hline
\end{tabular}
\tablebib{ $a$: \citet{Izotov:21b}; $b$: \citet{MendezDelgado:21a}; $c$: \citet{Esteban:13}; $d$: \citet{GarciaRojas:04}; $e$: \citet{Esteban:09}; $f$: \citet{DominguezGuzman:22}; $g$: \citet{Rodriguez:05}; $h$ \citet{Izotov:06}.}
\end{table*}

In Table~\ref{table:total_Fe_direct}, we present the ionic and total abundances in a sample of nebulae with detections of \feiv~$\lambda 6740$, which are useful for the direct estimation of Fe$^{3+}$/H$^{+}$ abundance. Other ionization states such as Fe$^{+}$ and Fe$^{4+}$ can be estimated using the lines \feii~$\lambda 8617, 8892$ \citep{Mendoza:23} and \fev~$\lambda 4227$. To calculate the abundances of Fe$^{+}$, we adopted the same temperature used to estimate the abundances of O$^{+}$, N$^{+}$, and Fe$^{2+}$, while for the abundances of Fe$^{3+}$ and Fe$^{4+}$, we adopted the same temperature used for O$^{2+}$ (see Sec.~\ref{sec:physical_chemical}). Note that the contributions of Fe$^{+}$ and Fe$^{4+}$ to the total Fe abundances are generally very small, while the contribution of Fe$^{3+}$ is always significant. The direct sum of ionic abundances can be compared directly with the predictions of Fe-ICFs by \citet{Rodriguez:05} and \citet{Izotov:06}, based on the predictions of photoionization models using the measured abundance of Fe$^{2+}$ and the degree of ionization. This exercise is performed for both $t^2=0$ and $t^2>0$. In this last case, the ionic abundances of Fe$^{3+}$ and Fe$^{4+}$ increase \citep{MendezDelgado:23a}, as well as the degree of ionization O$^{2+}$/O$^{+}$, which modifies the predictions of the Fe-ICF models.

\begin{figure}[h]
\centering    
\includegraphics[width=\hsize ]{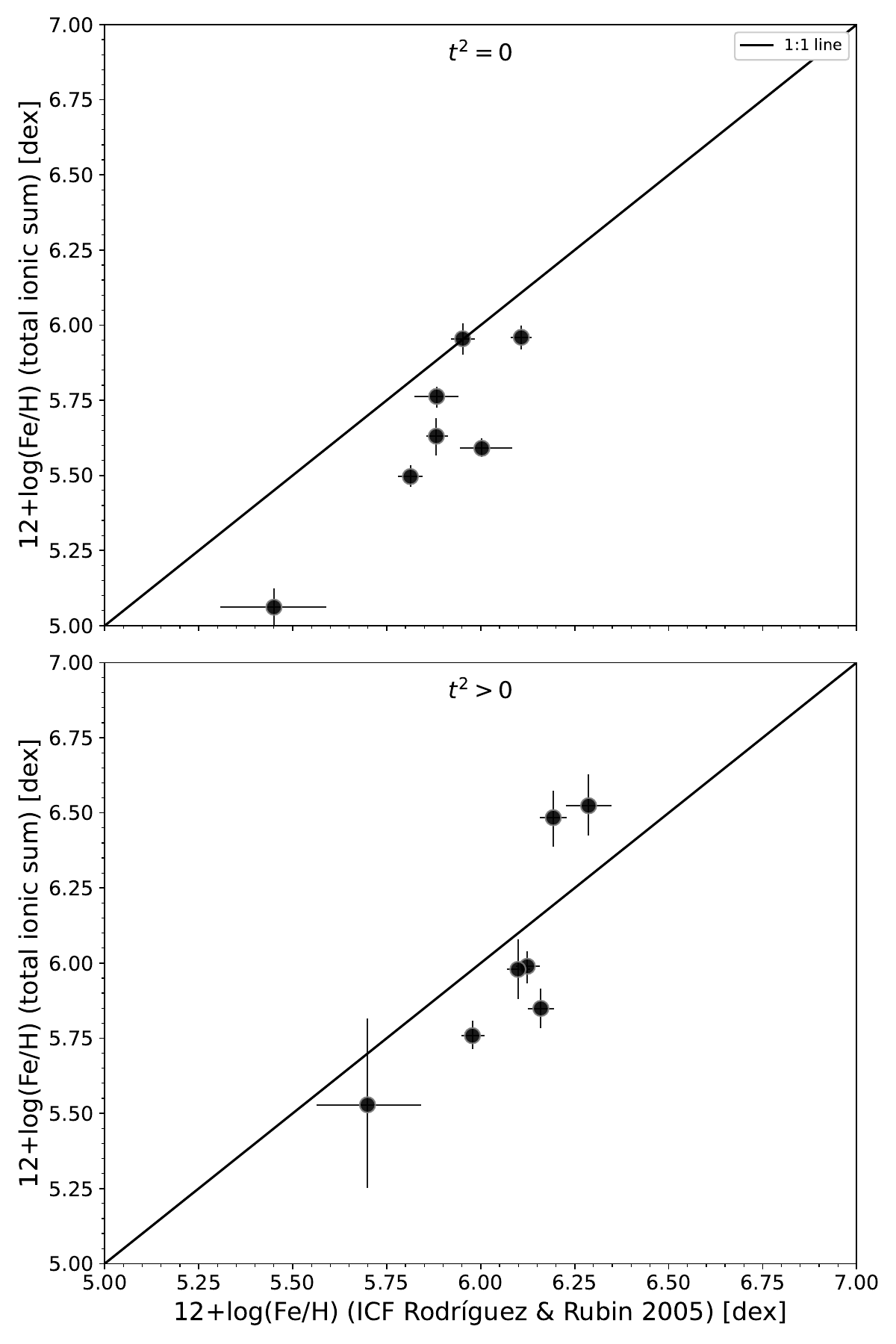}
\caption{Comparison between the Fe abundances obtained directly from the sum of the ionic abundances present in the gas in the sample of regions of the Table~\ref{table:total_Fe_direct} and those determined using the Fe-ICF of \citet{Rodriguez:05}, derived from photoionization models. Upper panel: the chemical abundances were calculated considering a homogeneous nebular temperature structure ($t^2=0$). Lower panel: the chemical abundances were calculated considering the presence of temperature inhomogeneities ($t^2>0$).
\\
} 
\label{fig:ICF_Rodriguez}
\end{figure}

\begin{figure}[h]
\centering    
\includegraphics[width=\hsize ]{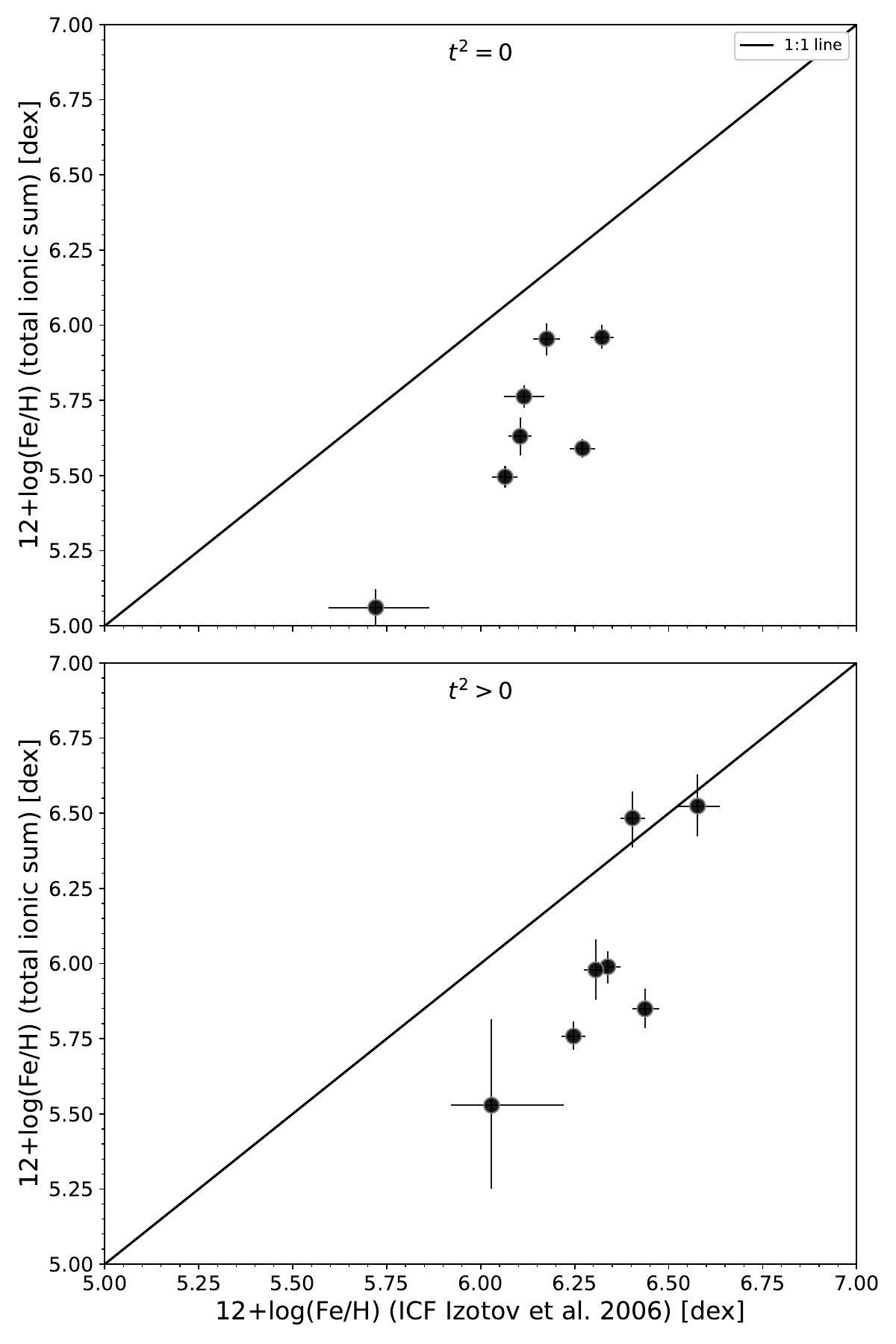}
\caption{Figure analogous to Fig.~\ref{fig:ICF_Rodriguez} but considering the Fe-ICF scheme of \citet{Izotov:06}.} 
\label{fig:ICF_IZOTOV}
\end{figure}

As shown in Figures \ref{fig:ICF_Rodriguez} and \ref{fig:ICF_IZOTOV}, both Fe-ICFs exhibit a high dispersion compared to the directly determined values. Notably, the Fe-ICF by \citet{Izotov:06} overestimates the total Fe abundances in almost all cases, showing differences of up to $\sim 0.7$ dex compared to the directly estimated total value. Dis\-cre\-pan\-cies bet\-ween the predictions of photoionization models and the directly determined gaseous Fe values, considering $t^2=0$, have been known since the pioneering work of \citet{Rubin:97}. This discrepancy, known as ``The \feiv~discrepancy'' \citep{Rodriguez:05}, consists of photoionization models predicting Fe$^{3+}$ abundances higher than those obtained observationally through the direct method, considering $t^2=0$. This problem is clearly seen in the upper panels of Figures \ref{fig:ICF_Rodriguez} and \ref{fig:ICF_IZOTOV}, where practically all objects show higher Fe abundances in calculations using ICF. This issue appears to improve when considering the effects of temperature inhomogeneities, using the formalism of \citet{Peimbert:67} and the empirical results of \citet{MendezDelgado:23a}. Under this scheme, empirical abundances of highly ionized ions, such as Fe$^{3+}$, increase when correcting the temperature bias introduced by an inhomogeneous physical conditions structure \citep{Peimbert:67,Cameron:23}. %The photoionization models used by \citet{Rodriguez:05} and \citet{Izotov:06} were created under conditions of homogeneous density and temperature, which may not be realistic in empirical determinations that

Although in the lower panel of Fig.~\ref{fig:ICF_Rodriguez}, the data points lie both above and below the line (suggesting statistical rather than systematic errors), the dispersion remains very high at around $\sim 0.2$ dex. This dispersion may be due to various factors, including errors in the atomic models of Fe, as suggested by \citet{Rodriguez:05}, or to the  assumptions made to determine the ionic abundances. However, it is important to mention the possibility that such dispersion may be real, resulting from various physical phenomena. We suggest the possibility that the fraction of Fe trapped in dust may differ between the volume where Fe$^{2+}$ and Fe$^{3+}$ are present. It is expected that the energy of photons interacting with dust grains is different in the low and high ionization volumes, inducing a different dust grain fragmentation rate. Additionally, shock waves and stellar winds in the different ionization volumes may play a role. If the fraction of Fe trapped in dust grains is higher in the volume where Fe$^{2+}$ coexists than where Fe$^{3+}$ does, then an ICF based on the gaseous fraction of Fe$^{2+}$ and the degree of ionization will underestimate the gaseous abundance of Fe, and vice versa. It is also possible that radiation pressure and stellar winds are capable of moving dust grains \citep{Rodriguez:02} from one ionization volume to another, inducing concentration variations in the nebulae.

Considering the present discussion, we adopt as default the ICF scheme from Eq.~2 of \citet{Rodriguez:05} and the nebular abundances considering $t^2>0$. However, a greater number of nebulae with detections of the \feiv~$\lambda 6740$ line or another transition of that ion would be highly beneficial for constraining potential errors in the total estimation of gaseous Fe in ionized nebulae.

\section{Gas phase Fe distributions}
\label{sec:results}

\subsection{Fe and O nebular abundances}
\label{subsec:FevsOdis}

\begin{figure}[h]
\centering    
\includegraphics[width=\hsize ]{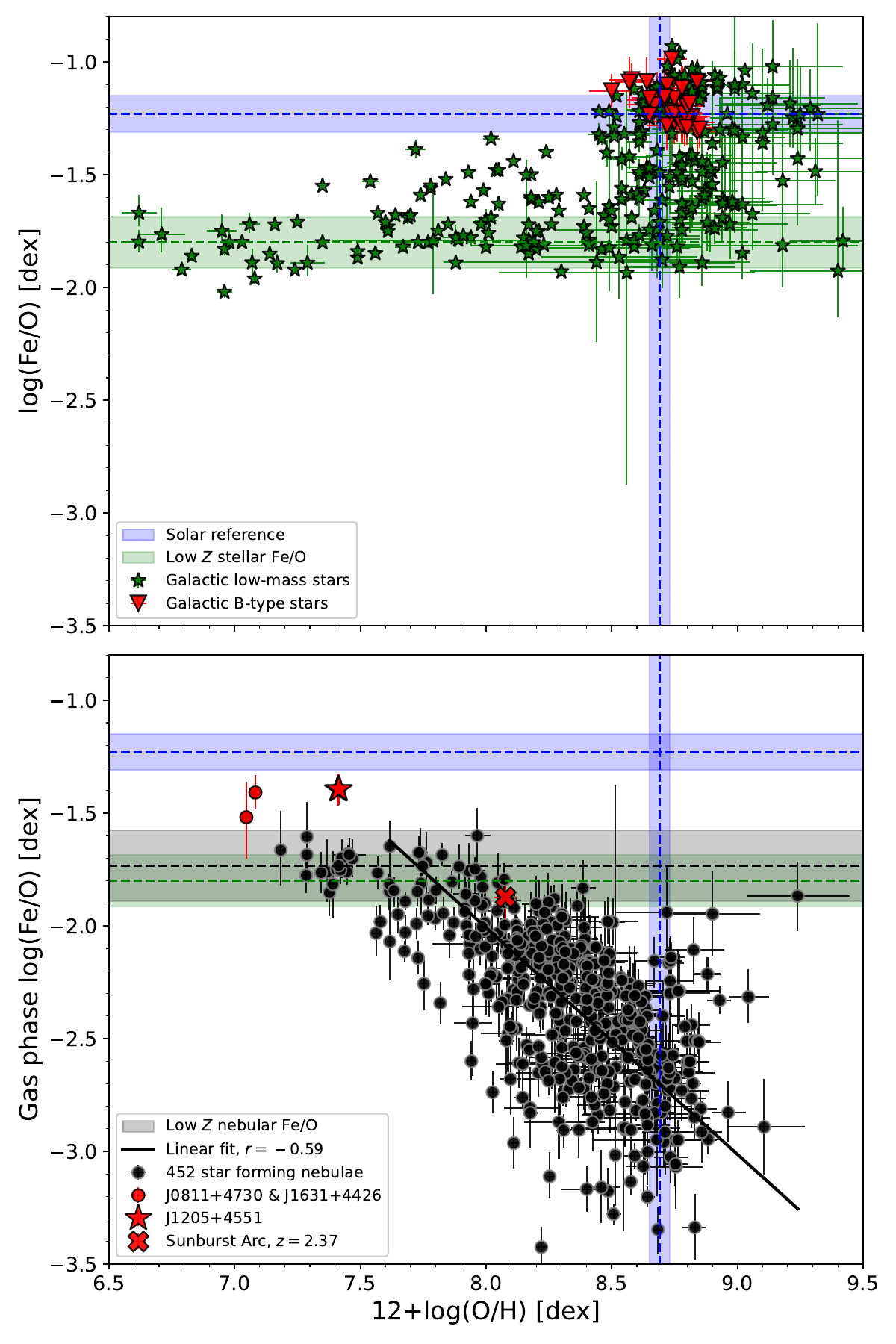}
\caption{Distribution of log(Fe/O) vs 12+log(O/H). Upper panel: Milky Way stars from the sample described in Sec.~\ref{sec:obs}. Lower panel: Galactic and extragalactic star-forming nebulae. The red dots highlight the position of J0811+4730 \citep{Izotov:18a} and J1631+4426 \citet{Kojima:21}, recently interpreted by \citet{Kojima:21} as enriched in Fe. The red stars indicate the position of J1205+4551 \citep{Izotov:17b, Izotov:21b} a galaxy with elevated Fe/O abundances and evidence of WR activity. The red cross points out the location of the Sunburst Arc \citep{Welch:24}, a high-redshift ($z=2.37$) galaxy recently observed by the JWST. Nebular O/H estimates have considered the influence of nebular temperature variations ($t^2>0$). The solar Fe/O and O/H abundances from \citet{Asplund:21} are shown as a reference. $r$ is the Pearson correlation coefficient of the linear fit.
}
\label{fig:resulting_FeO}
\end{figure}

In Fig~\ref{fig:resulting_FeO}, we present the distribution of Fe/O, both in Galactic stellar objects and in Galactic and extragalactic star-forming nebulae. In the stellar panel, the observed distribution is well-known \citep{Amarsi:19, Kobayashi:20, Chruslinska:23} and exhibits an increasing trend with respect to O/H. This is consistent with the predictions from the different enrichment timescales of CCSNe as primary O producers and SNe-Ia as primary Fe producers \citep[e.g.,][]{Chruslinska:23}. It is important to note that at the low metallicities covered in our sample, Fe/O remains relatively constant, with log(Fe/O)=$-1.80\pm 0.11$, consistent with the value reported by \citet{Amarsi:19} and \citet{Chruslinska:23} for Galactic low-metallicity dwarf stars. The stellar distribution of Fe/O vs O/H resembles the distribution of N/O versus O/H observed in nebular regions \citep{Henry:00, Nava:06, Nicholls:17}. These analogous behaviors can be explained by the similar timescales required to form a white dwarf in a stellar system that gives rise to a SN-Ia (Fe producers) and that required for an intermediate-mass star to release the N produced via CNO processes ($\sim 40$ Myr).

Although there is a relatively small number of objects (23, see Table~\ref{table:equations_Fe_O_N}) with 12+log(O/H)<7.6, most of the star-forming nebulae in the lower panel of Fig.~\ref{fig:resulting_FeO} seem to exhibit a flattening in the Fe/O distribution around log(Fe/O)=$-1.74 \pm 0.15$, rather consistent within the errors with the log(Fe/O)=$-1.80\pm 0.11$ determined in stellar objects. This may indicate that regions within this range of O/H abundances exhibit a negligible fraction of Fe depleted in dust. Under these conditions, the gaseous abundance of Fe would be representative of the total abundance of this element in ionized nebulae.

In the lower panel of Fig.~\ref{fig:resulting_FeO}, as the O/H abundance increases beyond 12+log(O/H)$\approx$7.6, Fe/O rapidly decreases, in contrast to the behavior observed in stellar objects. This trend is the result of Fe depletion onto dust grains within the ionized nebulae \citep{Rodriguez:02, Rodriguez:05, Izotov:06, Izotov:21b}. The fraction of Fe trapped in dust grains increases with metallicity mainly due to two factors. First, as metallicity increases, the ionizing sources in the nebulae tend to be softer \citep{Vilchez:88a, Stasinska:15}. This makes it less likely for dust grains in the ionized gas to shatter upon being hit by energetic photons \citep{Rodriguez:02}. Secondly, the formation of solid Fe compounds is proportional to the availability of this element. As seen in the upper panel of Fig.~\ref{fig:resulting_FeO}, the abundance of Fe/H increases more rapidly than O/H at high metallicities.

%\citep{Berg:13, Berg:20, Croxall:15, Croxall:16, DelgadoInglada:16, DominguezGuzman:22, Esteban:04, Esteban:09, Esteban:13, Esteban:14, Esteban:17, Esteban:18, Esteban:20, Fernandez:18, Fernandez:22, GarciaRojas:04, GarciaRojas:05, GarciaRojas:06, GarciaRojas:07, Guseva:09, Guseva:11, Hagele:06, Izotov:06, Izotov:09, Izotov:12, Izotov:17b, Izotov:21b, Kojima:21, LopezSanchez:07, MendezDelgado:21a, MendezDelgado:21b, MendezDelgado:22b, MesaDelgado:09, Peimbert:03, Peimbert:12, Rogers:22, Watanabe:24, Zurita:12}
\begin{figure}[h]
\centering    
\includegraphics[width=\hsize ]{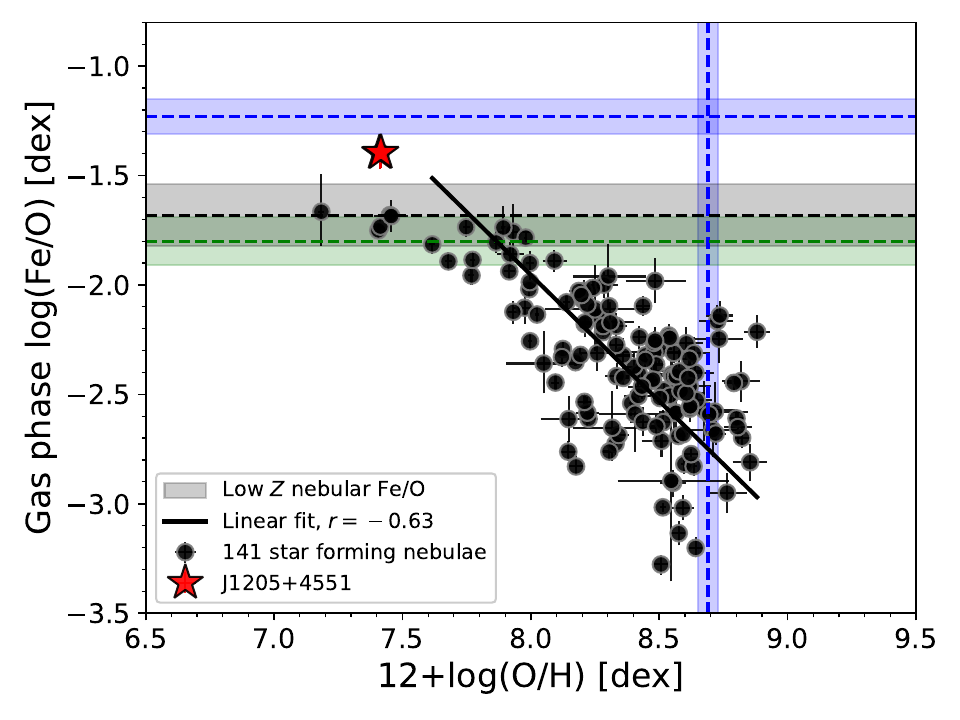}
\caption{Same as the lower panel of Fig.~\ref{fig:resulting_FeO} but considering only regions with errors in the \feiii~$\lambda 4658$ line intensity smaller than 10\%.} 
\label{fig:resulting_FeO_smallerrors}
\end{figure}

Although the decreasing trend of Fe/O at high values of O/H is clear, the dispersion is relatively high ($\sim 0.3$ dex) and the Pearson correlation coefficient is only moderate ($r=-0.59$). Based on Sloan Digital Sky Survey (SDSS) spectra \citep{Abazajian:05}, \citet{Izotov:06}  have suggested that the high dispersion in the Fe/O relation may be dominated by the errors in the flux measurements of the faint \feiii~lines. In Fig.~\ref{fig:resulting_FeO_smallerrors}, we consider the subsample of our nebular regions with the deepest spectra, with uncertainties in the \feiii~$\lambda 4658$ intensity below 10\%. This figure shows a correlation coefficient practically identical to that found in the general sample of nebulae and a relatively high dispersion, suggesting that much of the dispersion is real and caused by a physical reason as those mentioned in Sec.~\ref{sec:ICF_Fe_errors}.

An important factor that may contribute to the scatter is the presence and propagation of shocks  
   in the nebular gas \citep{Rodriguez:02}. Several studies \citep{Blagrave:06, MesaDelgado:09, Espiritu:17, MendezDelgado:22b} have observationally demonstrated the ability of photoionized shocks to break dust grains and release Fe into its gaseous phase. However, we propose that the relationship between Fe/O and O/H does not exhibit a higher linear correlation simply because the phenomena inducing Fe depletion in dust grains are not perfectly linearly related to the abundance of O/H. Although there is a relationship between the radiation hardness and the abundance of O/H, it is not linear \citep{Morisset:04, SimonDiaz:08}. In contrast, the effective temperature of the ionizing stars and the importance of their stellar winds depends primarily on Fe rather than O \citep{Garcia:14,Chruslinska:23}. More importantly, the total abundance of Fe, on which the formation of Fe-rich dust compounds could depends, does not scale linearly with that of O at high metallicity, as shown in the upper panel of Fig.~\ref{fig:resulting_FeO}.

\subsection{Fe and N nebular abundances}
\label{subsec:FevsNdis}

\begin{figure}[h]
\centering    
\includegraphics[width=\hsize ]{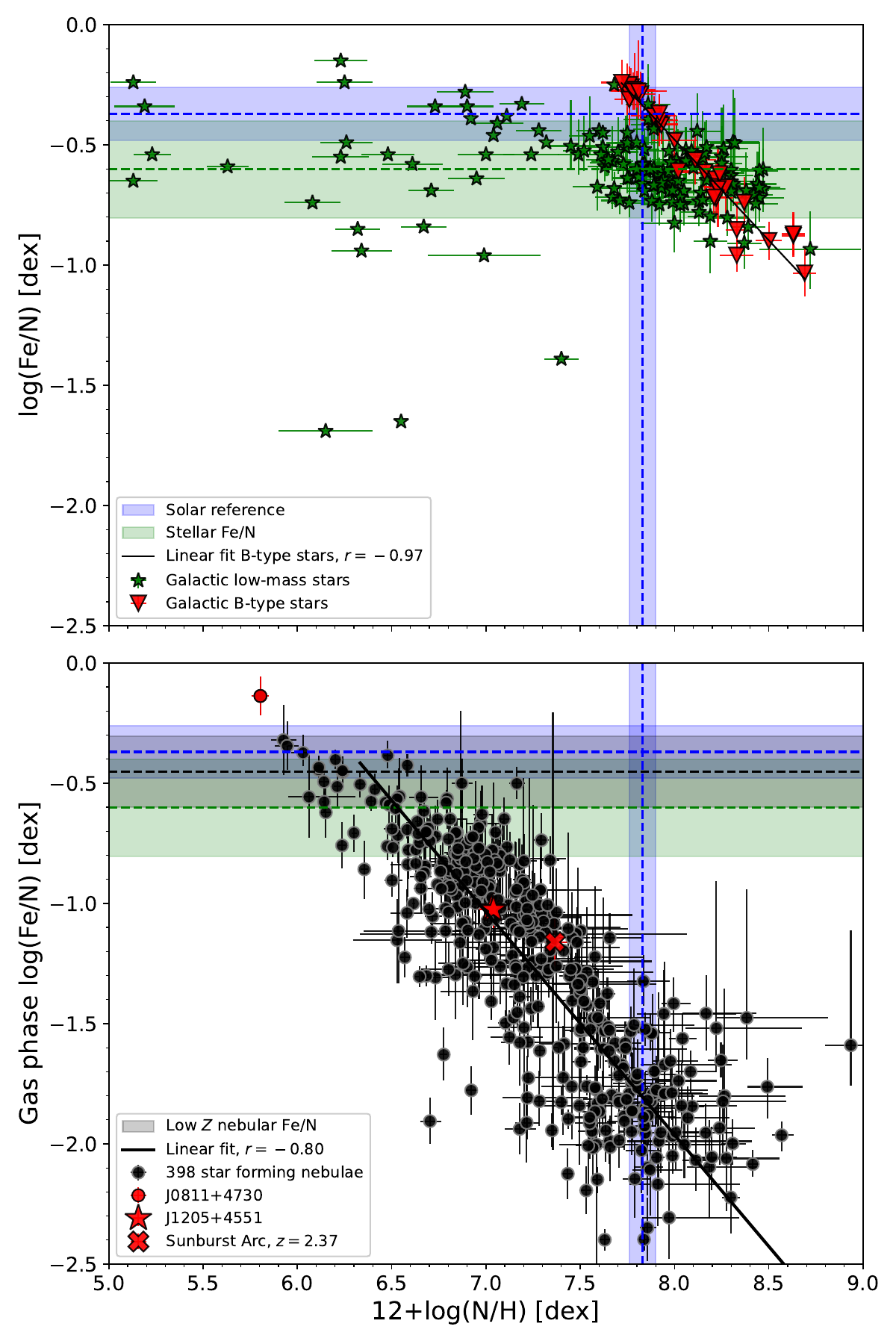}
\caption{Distribution of log(Fe/N) vs 12+log(N/H). Upper panel: Milky Way stars from the sample described in Sec.~\ref{sec:obs}. Lower panel: Galactic and extragalactic star-forming nebulae. The red dot highlights the position of J0811+4730 \citep{Izotov:18a}, recently interpreted by \citet{Kojima:21} as enriched in Fe. The red star indicates the position of J1205+4551 \citep{Izotov:17b, Izotov:21b} a galaxy with elevated Fe/O abundances and evidence of WR activity. The red cross points  out the location of the Sunburst Arc \citep{Welch:24}, a high-redshift ($z=2.37$) galaxy recently observed by the JWST. Nebular N/H estimates have considered the influence of nebular temperature variations ($t^2>0$). The solar Fe/N and N/H abundances from \citet{Asplund:21} are shown as a reference. $r$ is the Pearson correlation coefficient of the linear fit.
} 
\label{fig:resulting_FeN}
\end{figure}

Considering the hypothesis that the timescale to form a white dwarf in a stellar system that produces a SN-Ia is similar to the timescale over which an intermediate mass star returns secondary N to the ISM, we decided to explore the Fe/N distribution. If this hypothesis holds true, then the production of Fe at high metallicities should be better correlated with that of N rather than O. In Fig.~\ref{fig:resulting_FeN}, we present the distribution of log(Fe/N) as a function of 12+log(N/H) both in Milky Way stars (upper panel) and in star-forming nebulae (lower panel), analogous to what is presented in Fig.~\ref{fig:resulting_FeO}.

%Note: No Fe II above 18kK (B2), no Fe III above ~24kK (B0.5 spectral type)

Despite the significant dispersion in the upper pannel of Fig.~\ref{fig:resulting_FeN}, the observed Fe/N ratio in low-mass stars can mostly be encompassed around a relatively flat band, represented by a green dashed line in Fig.~\ref{fig:resulting_FeN} and which corresponding parameters are given in Table \ref{table:equations_Fe_O_N}. This trend can be explained by the findings of different studies where the stellar abundance of Fe scales linearly with that of N even in wider abundance ranges than those considered in this work \citep{Israelian:04,Ecuvillon:04,Magrini:18,Kobayashi:20, Grisoni:21}. Notably, when examining the B-type stars studied by \citet{Nieva:12} and \citet{Wessmayer:2022}, shown in red triangles in the upper pannel of Fig.~\ref{fig:resulting_FeN}, a strongly linear Fe/N correlation is observed, contrasting with the Fe/O relationship observed in this same stellar sample. These stars have formed more recently than the lower-mass star sample and have been enriched with metals. It is likely that the N present in these B-type stars in the solar neighborhood has been modified via the CNO cycle and mixing processes \citep{Przybilla:10}. In such case, these stellar N abundances would not be comparable to those of \hii~regions or other stellar systems. On the other hand, the Fe/H abundance shows little variation, suggesting homogeneous production of Fe/H in the interstellar medium that contributed to the formation of these B stars. This will be further discussed in Sec.~\ref{subsec:stars_vs_nebulae}.

\begin{table*}
    \caption{Fitted distributions of the Fe/O and Fe/N abundances shown in Figures \ref{fig:resulting_FeO} and \ref{fig:resulting_FeN}.}
    \label{table:equations_Fe_O_N}
    \begin{tabular}{cccccc}
        \hline
        \noalign{\smallskip}
        Abundance ratio & Relation & Range & N &Pearson $r$&$\sigma$ [dex]\\
        \noalign{\smallskip}
        \hline
        \noalign{\smallskip}
        \multicolumn{6}{c}{Star-forming nebulae}\\
        \hline
        \multirow{2}{*}{log(Fe/O)}&$-1.74 \pm 0.15$ & $12+\text{log(O/H)} < 7.6$&23&-&\multirow{2}{*}{0.29}\\
        &$ (-1.00 \pm 0.05)\times\left[12+\text{log(O/H)}\right]+(6.03\pm 0.38) $ & $12+\text{log(O/H)} \geq 7.6$& 429&-0.59\\
\\
        \multirow{2}{*}{log(Fe/N)}&$-0.45 \pm 0.15$ & $12+\text{log(N/H)} < 6.3$&16&-&\multirow{2}{*}{0.28}\\
        &$ (-0.91 \pm 0.03)\times\left[12+\text{log(N/H)}\right]+(5.47\pm 0.21) $ & $12+\text{log(N/H)} \geq 6.3$& 382&-0.80\\   
        \hline
        \multicolumn{6}{c}{Low-mass stars}\\
        \hline
        \multirow{1}{*}{log(Fe/O)}&$-1.80 \pm 0.11$ & $12+\text{log(O/H)} < 7.6$&27&-&-\\
        \multirow{1}{*}{log(Fe/N)}&$-0.60 \pm 0.20$ & $5.0\leq 12+\text{log(N/H)} \leq 8.7$&143&-&-\\
        \hline
        \multicolumn{6}{c}{B-type stars}\\
        \hline
        \multirow{1}{*}{log(Fe/N)}&$ (-0.84 \pm 0.04)\times\left[12+\text{log(N/H)}\right]+(6.25\pm 0.34) $ & $ 7.7 \leq 12+\text{log(N/H)} \leq 8.7$& 34&-0.97&0.06\\   

        \noalign{\smallskip}
        \hline
    \end{tabular}
\end{table*}

In the lower panel of Fig.~\ref{fig:resulting_FeN}, we show the distribution of gaseous Fe/N in star-forming nebulae, similar to that shown in the lower panel of Fig.~\ref{fig:resulting_FeO} for Fe/O. The higher Pearson correlation coefficient suggests that the linear correlation between Fe/N abundance and N/H is stronger than that observed for Fe/O and O/H, althought the dispersion remains quite high $\sim 0.3$ dex, similar to what is found in the Fe/O vs O/H distribution. This could indicate that some of the key factors in the Fe dust depletion have a closer relationship with N abundance than with O. Given the close relationship between the stellar abundances of Fe and N observed in the upper panel of Fig.~\ref{fig:resulting_FeN}, it seems plausible that the nucleosynthetic production of Fe is better correlated with that of N, due to similarities in the timescales required for SN-Ia production and the evolution of intermediate-mass stars. The use of N as a proxy for Fe would allow us to capture the systematic effect of radiation hardness, which is capable of destroying dust grains in photoionized environments \citep{Rodriguez:02} and scales with the abundance of Fe/H \citep{Garcia:14, Chruslinska:23}. Additionally, N exhibits insignificant depletions even in neutral environments of high metallicities \citep{Jenkins:09}. \\

\begin{figure}[h]
\centering    
\includegraphics[width=\hsize ]{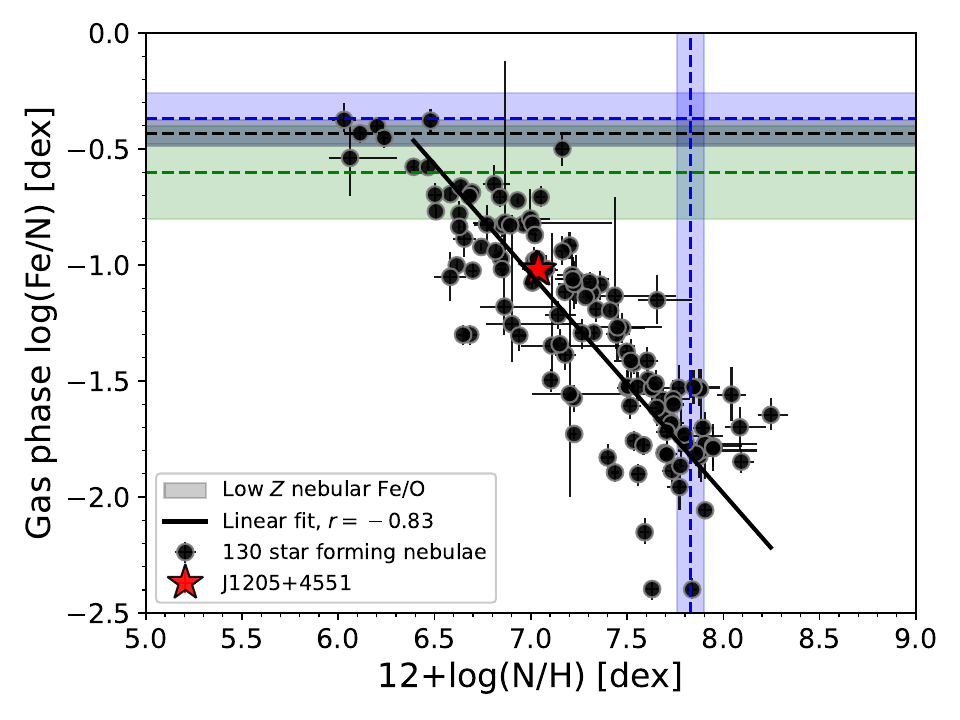}
\caption{Same as the lower panel of Fig.~\ref{fig:resulting_FeN} but considering only regions with errors in the \feiii~$\lambda 4658$ line intensity smaller than 10\%.}
\label{fig:resulting_FeN_smallerrors}
\end{figure}

Similarly to what is presented in Fig.~\ref{fig:resulting_FeO_smallerrors} for the Fe/O vs. O/H distribution, in Fig.~\ref{fig:resulting_FeN_smallerrors} we show the Fe/N vs. N/H distribution considering only the regions where \feiii~$\lambda 4658$ has uncertainties in its intensity lower than 10\%. This figure shows that the general distribution observed in Fig.~\ref{fig:resulting_FeN} is maintained, although with less dispersion. Additionally, an apparent flattening in the distribution is observed when 12+log(N/H) < 6.3. This flattening is not observed in the general sample and will be discussed in more detail in Sec.~\ref{sec:discussion}.

\section{Discussion}
\label{sec:discussion}

\subsection{Possible empirical relations for dust depletion}
\label{subsec:empirical_dust}

As shown in Fig.~\ref{fig:resulting_FeO} when 12+log(O/H) $<$ 7.6 the nebular gas-phase Fe/O seems to reach a constant value, or at least change the slope of the general trend, being consistent with the abundance one observed in Galactic stars in the same metallicity range. This flattening may suggests that the fraction of Fe trapped into dust in these photoionized nebulae is very small. This could be explained by the fact that the photoionization conditions in metal-poor regions are harder \citep{Vilchez:88a, Stasinska:15}, making the destruction of dust grains more efficient. At the same time, the availability of Fe, essential for the formation of some solid compounds, is very low. Determining the metallicity ranges where total Fe can be directly inferred from gaseous Fe is very important because it could point out to significant differences in the impact of dust in local ionized environments (generally of higher metallicity) and those observed in less evolved galaxies, such as those currently detected with the JWST. This could also help to explain why some Fe lines have been detected in high-z galaxies \citep{ArellanoCordova:22, Ji:2024, Tacchella:24, Welch:24}, despite being very weak in local \hii~regions.

However, the flattening effect is not clearly seen in the general distribution of Fe/N presented in Fig~\ref{fig:resulting_FeN}. This is a bit puzzling as it seems well established that nebular N/O abundance reaches a plateau at low metallicities \citep{Garnett:90, Nava:06, Skillman:13, Nicholls:17} and therefore, one would also expect to observe a flattening in Fe/N if it is present in Fe/O. In our sample, the regions 0556-51991-31, J0811+4730, and SBS-0335-052E \citep{Izotov:06, Izotov:09, Izotov:18a} seem to extend the linear trend between Fe/N and N/H observed at high metallicities. This could be a consequence of the changes in the gaseous abundance of Fe due to different depletion patterns being small and going unnoticed when compared to the abundance of O, as it is an element much more abundant than N. Alternatively, this could simply be a problem of low statistics when 12+log(N/H) $<$ 6.5. In fact, such flattening seems to begin to appear in Fig.~\ref{fig:resulting_FeN_smallerrors}, which shows the Fe/N distribution only in regions with the best signal-to-noise ratio in \feiii$\lambda 4658$. Notably, in low metallicity regions, the detection of \nii~$\lambda \lambda 6548, 6584$ is more challenging than in the case of \feiii~$\lambda 4658$. In fact, in this work, we have more regions with direct determinations of Fe and O than that of Fe and N. In our DESIRED sample, using the same criteria described in Sec.~\ref{sec:physical_chemical}, we determine that when 12+log(O/H) $<$ 8.0, log(N/O) $\sim-1.3$ (Arellano-Córdova et al. in prep.). Therefore, we adopt 12+log(N/H) = 6.3 as equivalent to 12+log(O/H) = 7.6. However, we find a relatively high dispersion ($\sim0.3$ dex) around this value.

If we assume that the total abundance of Fe scales proportionally to that of N, as suggested by stellar abundances and the previous discussion, it is possible to determine an empirical relationship to approximately determine the fraction of Fe trapped in dust as a function of the N/H abundance. In Eq.~\eqref{eq:Fe_Dustratio_derived_N}, we present this relation considering only the nebular values given in Table \ref{table:equations_Fe_O_N}. Eq.~\eqref{eq:Fe_Dustratio_derived_N} is valid for high-metallicity nebulae, where 12+log(O/H) $\geq$ 7.6 or 12+log(N/H) $\geq$ 6.3. For values lower than these,  $\text{Fe}_\text{Dust}/\text{Fe}_\text{Total} \approx 0$. 

\begin{equation}
\label{eq:Fe_Dustratio_derived_N}
\frac{\text{Fe}_{\text{Dust}}}{\text{Fe}_{\text{Total}}} \approx 1-1\times 10^{-5} \times \left( \frac{\text{N}}{\text{H}}\right)^{-0.91} . 
\end{equation}

\begin{figure*}
\centering    
\sidecaption
\includegraphics[width=12cm ]{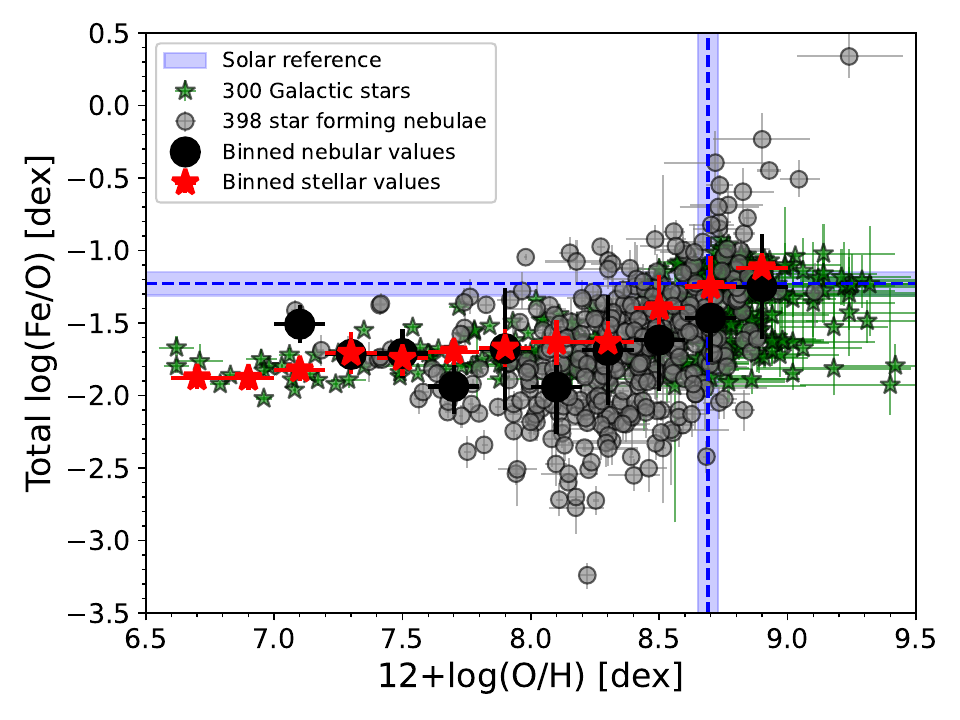}
\caption{Total Fe/O vs O/H abundance ratios for Galactic stars (green stars) and gas-phase abundance ratios for star-forming nebulae (gray circles). The plotted gas-phase abundances include the contribution from the dust-depleted Fe fraction calculated using Eq.~\eqref{eq:Fe_Dustratio_derived_N}. The binned values (with bins of $0.2$ dex) appear as red stars and black circles for Galactic stars and star-forming nebulae, respectively.}
\label{fig:Total_result}
\end{figure*}

Eq.~\eqref{eq:Fe_Dustratio_derived_N} implies that the formation of Fe-rich dust grains grows predominantly in proportion to the same Fe abundance. From this equation, the proportionality factor ($1 \times 10^{-5}$) presents the greatest uncertainties as it is mainly based on the Fe/N nebular values when 12+log(N/H) $<$ 6.3. In contrast, the power ($-0.91$) is much more robust as it represents the slope of the relationship between Fe/N and N/H considering 382 objects with 12+log(N/H) $>$ 6.3. Considering a solar N abundance of 12+log(N/H) = 7.83  \citep{Asplund:21}, Eq.~\eqref{eq:Fe_Dustratio_derived_N} indicates that approximately $\sim 95$\% of the Fe in ionized nebulae is depleted into dust grains. On the other hand, considering the values of the Large Magellanic Cloud (LMC) and the Small Magellanic Cloud (SMC) (12+log(N/H)=7.1-7.2 and 12+log(N/H)=6.5-6.7, respectively, see  Table~\ref{table:total_abundances}), the fractions would be approximately $\sim75$\% and $\sim35$\%, respectively. In the case of the SMC, the nitrogen abundance values are at the limit of validity of Eq.~\eqref{eq:Fe_Dustratio_derived_N}. The fraction of $\sim 35$\% assumes a value of 12+log(N/H)=6.7, but a negligible O-dust fraction of $\sim 1$\% is expected if 12+log(N/H)=6.5 is considered. In Fig.~\ref{fig:Total_result}, we show the total Fe/O abundance in the entire sample of star-forming nebulae, considering the gaseous fraction of Fe/H, measured from the \feiii~emission, and the fraction of Fe/H depleted into dust, inferred from Eq.~\eqref{eq:Fe_Dustratio_derived_N}. There is a rather good consistency between the stellar and nebular abundance distributions, as indicated by the binned values marked with red stars and black dots. 

%The good consistency in both stellar and nebular objects indicates that our nebulae in each range of O/H abundances are presenting similar stellar formation rates to those present in the ISM that gave rise to the stars in the past. It is well known that the O/Fe ratio is indicative of the stellar formation history of a system \citep{Kobayashi:06,Magrini:10, Amarsi:19,Chruslinska:23}.

Typically, the Fe/O vs. O/H relationship (or more commonly [O/Fe] vs. [Fe/H]\footnote{when dealing with abundances from stellar atmospheres, the typical notation is [X/Y]$=\log(N_{\rm X}/N_{\rm Y}) - \log(N_{\rm X}/N_{\rm Y})_\odot$, where $\log(N_{\rm X}/N_{\rm Y})_\odot$ is the number ratio measured in the Sun.}) is used to understand the star formation history of a particular galaxy \citep{Kobayashi:06,Magrini:10,Matteucci:12}. In this relationship, star formation efficiency and the Initial Mass Function are parameters that play a fundamental role. Similarly, these same parameters govern the N/O vs. O/H relationship \citep{Molla:06}. Considering a large number of star-forming regions from different galaxies, it is well established observationally that the N/O vs. O/H relationship reaches a plateau at low metallicities (12+log(O/H) $<$ 8) \citep{Henry:00,Nava:06,Nicholls:17}  determined by the onset of N production from intermediate mass stars \citep{Vincenzo:16}. If there is a close relationship between the production times of Fe and N, one would also expect a plateau in the Fe/O vs. O/H relationship among various star-forming regions, once the fraction of Fe trapped in dust grains has been considered. By looking at the binned nebular values in Fig.~\ref{fig:Total_result} we indeed observe a plateau up to roughly 12+log(O/H) = 8.0, with a subsequent rise in the Fe/O ratio. Moreover, the position of this ``knee", which indicates the onset of SNe Ia contribution, also agrees with the one observed in Galactic stars considered in this work. 
In turn, this means that, on average, star-forming regions in present-day local galaxies share a similar level of efficiency of star formation and Initial Mass Function shape relative to the one already experienced by the Galactic disk, capturing different evolutionary snapshots according to their metallicity. This appears to be at odds with stellar abundances observed in dwarf galaxies of the Local Group. For example, dwarf spheroidal galaxies show an O/Fe plateau (deduced from other $\alpha$-elements) roughly from 12+log(O/H) $<$ 7.5 \citep{Hendricks:14, Hill:19}. %These types of galaxies usually exhibit the O/Fe plateau at lower metallicities than other types of systems \citep{Tolstoy:09}. 
 However, these galaxies are faint, gas-poor, and quiescent in star formation (see e.g. \citealt{Tolstoy:09}). This indicates that nebular observations of galaxies with low O/H abundance will be biased toward regions of intense star formation corresponding to gas-rich, dwarf irregular galaxies. The position of the O/Fe plateau in irregular dwarf galaxies is still relatively uncertain, as there are very few determinations beyond local systems as the SMC and LMC, which might differ from the conditions of other lower-metallicity systems with a higher star formation rate.

Nonetheless, the position of the ``knee" could still show differences relative to the mean trend due to the different evolutionary history in each galaxy. This causes the significant dispersion in the nebular abundances in Fig.~\ref{fig:Total_result}, in addition to other factors contributing to the uncertainties in Fe, dominated by the correction for depletion onto dust grains. In fact, the distribution presented here in Fig.~\ref{fig:Total_result} is valid if the possible O/Fe plateau occurs when 12+log(O/H) $<$ 7.6. The precise position of the expected O/Fe-plateau may induce offsets in the nebular distribution of Fe/O observed, but in general, the functional form would not be altered. 

Fig.~\ref{fig:Total_result} also reinforces the fact that the conversion between stellar metallicities, measured with Fe, and nebular metallicities, measured with O, is not straightforward. Significant biases can be expected when transforming individual stellar Fe abundances to O abundances using the solar reference as in  \citet{Bresolin:09b, Bresolin:16, Bresolin:22}. In the best case scenario, when indeed we are within the range of solar metallicities, both stars and nebulae exhibit average dispersions of up to $\sim0.2$ dex in the Fe/O abundances, resulting from production variations between Fe and O. We think it is important to consider this uncertainty when comparing stellar and nebular metallicities. A much worse situation arises when comparing stellar and nebular galactic abundance gradients using the solar reference of Fe/O as a conversion factor as used in several works \citep[e.g.][]{Gazak:15, Bresolin:22}, as the entire abundance pattern changes with metallicity. Using the stellar reference of Fe/N that we present in Table~\ref{table:equations_Fe_O_N} could be potentially useful in this latter case. However, a larger statistical sample for stellar Fe and N abundances is required for this purpose. An alternative based on the knowledge of the specific SFR of the system is presented by \citet{Chruslinska:23}.

As suggested by \citet{Peimbert:10}, in the case that the depletion of O onto dust grains scales with that of Fe, it is possible to derive an upper limit to the O depletion. In the case of the Orion Nebula, it is known that the fraction of dust trapping O is up to $\sim 0.1$ dex \citep{MesaDelgado:09, Peimbert:10}. Therefore, it can be estimated that in the LMC and SMC there may be up to $\sim 0.07$ dex and $\sim 0.03$ dex of O trapped in dust, respectively. The depletion of O onto dust could be negligible at  12+log(O/H) $<$ 8.0, as its contribution is much smaller than the typical uncertainties in the estimation of chemical abundances in ionized nebulae. We remark that these could be considered upper limits to the depletion of O in photoionized environments as they are based on a scaling of the upper limit derived for the Orion Nebula. Therefore, simply adding 0.1 dex to correct for dust all nebular determinations of O/H, regardless of metallicity and ionization conditions --as it has been done in several works \citep[e.g.,][]{Bresolin:16, Bresolin:22}-- seems inappropriate.

\subsection{On the dust composition in \hii~regions}
\label{subsec:HII_dust}

The precise chemical composition of dust in ionized gas remains a matter of debate \citep{SimonDiaz:11}, and its impact on the important parameters of ionized nebulae could be significant \citep{Gunasekera:23}. The dust present in \hii~regions should consist of the grains most resistant to photoionization, formed prior to star formation. This is supported by the fact that we do not observe any connection between the depleted Fe fraction and the electron density, a case that could be expected if a significant fraction of the dust grains were formed within the photoionized environment \citep{Zhukovska:18}. In our Sec.~\ref{sec:appendix_CB}, we show a weak anticorrelation between optical reddening relative to H$\beta$ and the Fe/O and Fe/N abundances in a subsample of the regions analyzed in this work. The anticorrelation seems stronger in the case of Fe/N. This reinforces the idea of having an interconnection between the dust found in \hii~regions and that present in the neutral medium, which may cause most of the optical extinction. Studies of dust in the neutral ISM can provide candidates that may survive photoionization and be present in \hii~regions. In situ studies of dust in the neutral ISM indicate that it is likely that Fe is mostly locked up in free-flying iron particles and silicate grains \citep{Jones:17, Choban:22, Hensley:23, Dubois:24}.

Our results from Figs \ref{fig:resulting_FeO} and \ref{fig:resulting_FeN} show that the fraction of Fe trapped in dust grains decreases as metallicity decreases. Our interpretation is that photo-destruction is particularly important at lower metallicities \citep{Rodriguez:02, Rodriguez:05} because the ionizing spectrum becomes increasingly harder as the amount of Fe in stars decreases \citep{Vilchez:88a, Stasinska:15}. In turn, the gas-to-dust fraction in the neutral ISM tends to be higher at lower metallicities \citep{Gioannini:17, Galliano:21, Roman-Duval:22, Roman-Duval:22a}, which suggests that there is less Fe-rich dust in low-metallicity ionized environments because the fraction of Fe trapped in dust in the neutral ISM is lower prior the star formation. In fact, the relation we find between Fe/O and O/H is similar to the relation found by \citet{Gioannini:17} between Fe/Zn vs Zn/H and Fe/S vs S/H in neutral gas. Both the increased photo-destruction of dust grains in low-metallicity ionized environments and the reduced formation of dust in the low-metallicity neutral ISM are not mutually exclusive and could be acting in the same direction. It is important to mention that spatially resolved studies of the distribution of gaseous Fe/H in \hii~regions of our Galaxy, as is possible with the SDSS-V Local Volume Mapper \citep{Drory:24}, are of great importance to quantify the impact of dust photo-destruction in ionized environments. Preliminary results from the Orion Nebula (Méndez-Delgado et al. in prep) show an increasing trend of Fe/O and Fe/N when approaching the ionizing star $\theta^1$ Ori C, suggesting that the impact could be significant.

Photo-destruction should mainly affect the smaller grains, particularly the carbonaceous ones \citep{Jones:17}. Studies of photoionized Herbig-Haro objects in the Orion Nebula \citep{Blagrave:06,MesaDelgado:09,MendezDelgado:21a,MendezDelgado:21b,MendezDelgado:22b} show dramatic increases in the gaseous abundances of Fe, Ni and Cr at the bowshocks due to the destruction of dust present in the photoionized gas. In contrast, in these objects, the abundances of O and C remain virtually unchanged. This suggests that most of the Fe depleted in grains could be composed of Fe grains decoupled from C and O that could be efficiently destroyed by the ionizing radiation. This could suggest that most of the Fe-dust in ionized environments is composed by free-flying iron particles.

Interestingly, the fact that we find a significant depletion of Fe in dust grains in a large number of objects indicates that these grains were not yet processed by the supernova forward shock, which is predicted to destroy a large fraction of the dust grains \citep{Bocchio:14,Slavin:15,Kirchschlager:22, Kirchschlager:24}. However, given the large dispersion observed in the distributions of Fe/N and Fe/O, it cannot be ruled out that supernova shocks are playing a role in some fraction of the regions analyzed here.

\subsection{On the Fe enrichment on short timescales by very massive stars}
\label{subsec:very_massive}

Recently, \citet{Kojima:21} reported a puzzlingly high Fe/O in J0811+4730 and J1631+4426, two of the extremely metal-poor local dwarf galaxies in their sample. Such an abundance pattern cannot be explained by enrichment from regular CCSNe, and the inferred young age and very high specific SFR in those galaxies rule out the possibility of significant Fe enrichment by SNe Ia. \citet{Kojima:21} speculate that the high Fe/O could be indicative of enrichment by very massive stars with $M_\text{init} > 300M_{\odot}$. Other scenarios, involving enrichment by massive pair-instability supernovae and a non-universal IMF, have also been proposed \cite{Goswami:21}.

Our results presented in Sec.~\ref{sec:ICF_Fe_errors} suggest that \citet{Kojima:21} overestimate the Fe abundances in J0811+4730 and J1631+4426 due to the bias introduced by the Fe-ICF they use \citep[that proposed by][]{Izotov:06}. We argue that these abundances are overestimated by at least $\sim 0.2$ dex. However, we point out that the adoption of the ICF by \citet{Izotov:06} can introduce overestimations of the Fe abundances of up to $\sim 0.7$ dex in environments where the Fe$^{3+}$/Fe fraction is higher, which are precisely metal-poor regions like J0811+4730 and J1631+4426 \citep{Rodriguez:05}. 

%While we cannot rule out the possibility that the Fe/O values in these objects are elevated relative to the ``plateau" level (log(Fe/O) $\sim$ -1.7) occupied by other metal-poor regions, we do not see clear evidence for extreme early Fe enrichment in these systems either.

%This issue with the Fe-ICF is known since the work of \citet{Rubin:97} and is widely discussed by \citet{Rodriguez:05}. The so-called ``\feiv~discrepancy'', results from a systematic difference of the Fe$^{3+}$ abundance predicted by the photoionization models used to derive the ICF-scheme with respect to the direct estimations using $t^2=0$. This systematic bias can be more severe in environments where the Fe$^{3+}$/Fe fraction is higher, which are precisely metal-poor regions like J0811+4730 and J1631+4426. The use of $t^2>0$ with the ICF by \citet{Rodriguez:05} seems to provide better results as shown in the lower panel from Fig.~\ref{fig:ICF_Rodriguez}, but the dispersion remains high at $\sim 0.2$ dex.

%As shown in Fig.~\ref{fig:resulting_FeO}, J0811+4730 and J1631+4426 do not appear particularly anomalous compared to other regions with 12+log(O/H)<7.6 within the typical dispersion. 

Additionally, in contrast with the discussion presented by \citet{Kojima:21}, we show that the trends observed in Figures \ref{fig:resulting_FeO} and \ref{fig:resulting_FeN} can be explained in terms of the depletion of Fe onto dust grains in ionized environments. The existence of such dust is firmly established by various studies of local nebulae \citep{Rodriguez:99, Rodriguez:02, Rodriguez:05, Blagrave:06, MesaDelgado:09, MendezDelgado:21a} that include direct detections of the dust emissions \citep{Smith:05}. On the other hand, \citet{Kojima:21} did not find any correlation between the dust extinction and the Fe/O ratios. However, although local dust trapping Fe in Galactic nebulae could have some effects on the observed optical extinction due to their proximity to us \citep{Rodriguez:02}, this effect may be negligible in more distant systems where the influence of dust outside the \hii~region and integrated along the line of sight could play the most significant role.

Although J0811+4730 and J1631+4426 certainly exhibit a slightly elevated Fe/O abundance, they remain basically consistent with values obtained in other regions with low O/H abundances, explained in terms of preferential dust depletion. Therefore, we do not find compelling evidence of Fe overproduction by stars with $M_\text{init} > 300M_{\odot}$ in these objects.

Interestingly, in the lower panel of Fig.~\ref{fig:resulting_FeO} two spectra of J1205+4551 highlighted by the red stars \citep{Izotov:17b, Izotov:21b}, show Fe/O abundances similar to those found by \citet{Kojima:21} in J0811+4730 and J1631+4426. J1205+4551 additionally exhibits signatures of WR stars \citep{Izotov:21b} and a very high N/O abundance for its O/H abundance. In more massive stars, surface nitrogen quickly gets enriched at the expense of oxygen. Strong stellar winds can then lead to an enrichment of the ISM with nitrogen \citep{Meynet:2005, Crowther:07, LopezSanchez:07}. In particular the WN stage can provide an efficient channel here as the products of He burning have not yet reached the stellar surface, but the winds are stronger than for normal O stars. This effect is prominently seen in the abundances of resolved nebulae around massive WN stars \citep[e.g.,][]{Kwitter:84, Stock:11, Esteban:16}, and consequently predicted for the yields of very massive, hydro-burning stars ($M_\text{init} \geq 100\,M_\odot$) that also show WNh-type spectra \citep[e.g.,][]{Higgins:23}. The effect of WR stars on Fe is not entirely clear, mainly due to the high depletion of this element onto dust grains at high metallicities. 

In contrast to the Fe/O abundance, as shown in Fig.~\ref{fig:resulting_FeN}, the Fe/N abundance in J1205+4551 does not appear anomalous, but rather consistent with the general trend. Another interesting case is the Sunburst Arc, a galaxy at $z=2.37$ recently observed with the JWST and analyzed by \citet{Welch:24}. This latter galaxy presents a high N/O abundance, unexpected for its low O/H abundance. However, in Fig.~\ref{fig:resulting_FeN}, we find that, similarly to the case of J1205+4551, the Sunburst Arc (highlighted by the red cross) shows quite a normal Fe/N abundance. This might indicate that some nebular regions with abnormally high N/O abundances could also contain rather high Fe/O abundances. It is possible to speculate that the high values of Fe/O and N/O in these regions may be due to a lower availability of O resulting from the presence of WR stars. A systematic study of the Fe/H abundance in \hii~regions with low O/H abundances and high N/O ratios like some particular knots of NGC~5471 \citep{Kennicutt:03} or Mrk~71 \citep{Esteban:02}, may shed light on this interesting issue. 

\subsection{Stellar and nebular metallicities}
\label{subsec:stars_vs_nebulae}

The main idea to interpret the linear relation between gas-phase abundances of Fe and N in star-forming nebulae is that the production timescales of N and Fe are similar, while dust formation trapping Fe scales with its own availability. This seems valid in the range of chemical abundances studied here \citep{Magrini:18, Xiong:22, Sun:23}, and is consistent with the assumption of the primary origin of N in CCSNe of massive stars and secondary in intermediate-mass stars through the CNO cycle \citep{Henry:00, Meynet:02a, Meynet:02b}. In general, the different nucleosynthetic origins of N and Fe can add additional complexity to the empirical abundances of both elements. For instance, N enrichment from intermediate-mass stars is known to be sensitive to metallicity \citep{Ventura:13}. Such metallicity-sensitive yields could decouple the gas-phase Fe and N abundances, specially at supersolar metallicities. Nevertheless, within the metallicity range covered by this analysis, which encompasses most of the nebular analyses reported in the literature, such deviations are not observed. The yield predictions from \citet{Ventura:13} appear consistent with observational stellar values within the analyzed metallicity range, as illustrated in Figure~1 of \citet{Romano:19} (corresponding to models MWG06 and MWG07). Our study does not extend beyond the metallicity ranges discussed in these works. Future studies that cover metallicity ranges beyond those examined here could investigate these predictions.

Some authors have suggested the existence of massive stars with high rotation velocities as a source of N observed at low metallicities \citep{Limongi:18, Prantzos:18}, being able to successfully reproduce the observed flat trends between Fe/N and N/H in Galactic stars \citep{Prantzos:18, Limongi:18, Romano:19}. These differences in the precise origin of N at low metallicities do not seem to affect our interpretation, at least within the range of our chemical determinations, as the production timescales remain similar.

As observed in the upper panel of Fig.~\ref{fig:resulting_FeN}, Galactic B-type stars \citep{Nieva:12, Wessmayer:2022} show more significant variations in the Fe/N ratio compared to the other stars analyzed here. This trend is not limited to Galactic B-type stars but is also evident in other galaxies \citep{Trundle:05, Trundle:07, Evans:07, Hunter:07}. For instance, in the SMC, LMC, and NGC\,3109, massive B-type stars (both giants and dwarfs) analyzed by \citet{Hunter:07} and \citet{Evans:07} typically exhibit nitrogen overabundances of more than one order of magnitude compared to nebular values \citep{Pena:07, Toribio:17, DominguezGuzman:22}, reflecting that massive stars rapidly reach the so-called ``CN'' equilibrium as a part of the CNO cycle \citep{Przybilla:10}. Confirmed by recent efforts of the XShootU collaboration \citep{Martins:24}, it seems that in galaxies with lower overall O/H abundance, massive stars exhibit higher N/H overabundances. However, determining whether this is systematic or merely a selection bias towards the brightest stars in these systems goes beyond the scope of this study. Nevertheless, it is important to note that N/Fe and N/O ratios in these stars do not constitute the final value that will be released into the ISM after their death in CCSN.

Interestingly, the very high overabundances of N observed in some extragalactic B-type stars could significantly alter the abundances of C and O. If a significant portion of the N is produced at the expense of O, non-negligible decrements, on the order of $\sim 0.1-0.2$ dex, may be present \citep{Martins:24}. This complicates element-by-element comparisons between the abundances of C, N, and O among massive B-type stars and \hii~regions when high precision is required. Studies of individual, unevolved OB stars which do not exhibit significant mixing processes \citep[e.g.,][]{Rolleston:03,Ramachandran:21,Pauli:23,Martins:24} could provide a very useful testbed for the consistency between present-day stellar and nebular abundances. Yet, if achieving high precision in the comparisons between O abundances in stars and nebulae is already complicated, it becomes even more so when dealing with stellar metallicities derived from the analysis of combined absorption lines of various elements  \citep[but dominated by Fe-peak elements,][]{Urbaneja:08, U:09, Kudritzki:08, Kudritzki:12, Kudritzki:14, Kudritzki:16, Hosek:14, Patrick:15}, as done by \citet{Bresolin:16} in several galactic systems using the solar reference to transform Fe-metallicities into O-metallicities. Direct O measurements instead are typically focussed on very slow rotating stars such as the prototypical B-type dwarf \object{AzV~304} \citep{Rolleston:03}. While such stars allow for a high-precision O measurement due to the narrow lines, this limited selection might imply a bias in itself.

We are aware that the number of star-forming nebulae with 12+log(O/H) $<$ 7.6 in our sample is somewhat limited, but it already constitutes a significant observational achievement to have direct nebular determinations of Fe, O, and N under such conditions. The change in the slope of the Fe/O abundance ratio in this range of metallicities could be very small. If the value of log(Fe/O) $\sim$ $-$1.74 is representative of the total Fe/O abundances in these star-forming nebulae in the metallicity range --which correspond to dwarf irregular galaxies-- this could indicate similar star formation efficiencies than what was present, in the early stages of the Milky Way \citep{Carigi:05, Carigi:19, Nomoto:13, Amarsi:19, Kobayashi:06, Kobayashi:20, Prantzos:23}. In contrast, if new observations of star-forming nebulae were to show a continuation of the anti-correlation between Fe/O and O/H produced by Fe depletion in dust grains, this could indicate lower star formation efficiencies or the presence of some phenomenon such as metal losses through winds \citep{Tolstoy:09}. 

A high Fe/O ratio in low O/H environments would imply harder thermal equilibrium in ionized nebulae than what is typically observed in local \hii~regions. On the one hand, the stellar winds from ionizing stars are more significant as a function of the Fe abundance \citep[e.g.,][]{Vink:01,Garcia:14,Sander:20, Chruslinska:23}. On the other hand, gas cooling is mainly driven by the O abundance \citep{Osterbrock:06}. However, a potentially high rate of Fe production through SN-Ia events should occur on timescales similar to those of secondary N production in intermediate-mass stars, and it seems well established that in star-forming nebulae with 12+log(O/H) $<$ 8.0, the N/O abundance ratio is subsolar at around $\sim -0.5$ dex \citep{Henry:00, Nava:06, Pilyugin:10, Nicholls:17}, making this scenario less likely. Certainly, there may be specific star-forming nebulae with high N/O and Fe/O at low O/H due to their star formation history, similar to what is demonstrated by \citet{Molla:06} in the case of N/O. This could also contribute to the dispersion observed in the Fe/O ratios.

% \as{This trend is not limited to local B stars, but this should continue towards lower N/Fe if you add points from the LMC and SMC. I did a quick check and you should get up to Fe/N around 0.3 dex at log N/H + 12 about 7.0. Consider adding data from Korn+2000 and Trundle+2007. The same sources will give you an solar-like O/Fe for these B (dwarf) stars in the LMC and SMC, namely -1.13 for log O/H + 12 of 8.4 and -1.06 for 8.06. Moreover, you can get a few abundances for the two-three Fe-poor B stars in the Magellanic Bridge by combining Rolleston+1999 (CNO) and Dufton+2008 (Fe):, e.g. DGIK 975 with log N/H + 12 = 6.74 and Fe/N = -0.9 (the latter is probably a bit too low, but < -0.5 for sure). Similar for MBO3, a late O star from Ramanchadran+2020 with N/H + 12 = 6.68 where we now know from Elisa that Fe is ~0.05 solar and thus also Fe/O = -1.54 dex. Hence, only the Magellanic Bridge gives you B stars from different, "older" material. On a longer time scale, Elisa might be able to you more data points from B stars in the Magellanic Bridge, but this is likely too late for this paper. Note that this - the Bridge - includes partly unpublished material, so we should discuss what to use and cite.}

\subsection{Limitations of this work and future improvements}
\label{subsec:limitations}

One of the main objectives of the DESIRED project \citep{MendezDelgado:23b} is to unveil the chemical and physical properties of ionized nebulae that can be explored with the weakest emission lines, which are generally under-studied. In this case, the analysis of Fe abundances in the gaseous phase has critical limitations due to the estimation of the ICF, as detailed in Sec.~\ref{sec:ICF_Fe_errors}. The large dispersion of $\sim 0.2$ dex shown even in the best case  \citep[which uses $t^2>0$ and the ICF of][]{Rodriguez:05} substantially limits our ability to correctly interpret the Fe abundances in individual nebulae such as J0811+4730 and J1631+4426 \citep{Kojima:21}. Clearly, it is necessary to expand the sample of star-forming nebulae with measurements of \feiv~$\lambda 6740$ to refine the ICF. To our knowledge, such emission line has only been detected in 7 \hii~regions to date, which are analyzed here in Table~\ref{table:total_Fe_direct}. Given the significance of nebular Fe abundances, as evidenced in this article, allocating telescope time for the detection of weak \feiv~emission lines across a range of nebular systems is crucial.

Additionally, it is important to expand the detections of \feiii~in systems where 12+log(O/H) $<$ 7.6 to disentangle whether there is a plateau in the Fe/O ratios at such metallicities. The implications of a corroboration of this flattening, as well as those of a continuation of the anticorrelation between Fe/O and O/H, have been discussed in this article and are highly relevant for understanding the star formation history of galaxies. It is also important to understand if high Fe/O ratios are observed in star-forming nebulae with high N/O at low O/H, as in the case of J1205+4551 \citep{Izotov:17b, Izotov:21b}. This may potentially be occurring in some areas of NGC~5471 \citep{Kennicutt:03} or Mrk~71 \citep{Esteban:02} and could allow us to understand the role of WR stars in the chemical enrichment of the ISM. Finally, expanding the sample of Fe/N ratios in low-metallicity stars could allow us to establish a more robust link between the nucleosynthetic production of both elements.

\section{Conclusions}
\label{sec:concl}

Following the philosophy of the DESIRED project \citep{MendezDelgado:23b} of studying the nebular physics related to the weakest emission lines, we have analyzed the largest sample of star-forming regions with simultaneous direct determinations of the gas-phase Fe/H, O/H, and N/H ratios from the literature. We have determined the physical conditions ($n_e$, $T_e$) homogeneously and carefully and have considered the effects of temperature variations in the sample \citep[$t^2 > 0$,][]{Peimbert:67}, although our general conclusions are independent of the latter parameter.

We have shown a substantial discrepancy between the predictions of Ionization Correction Factor (ICF) models and the direct estimations of all ionization states of Fe. The greatest discrepancies were obtained with the ICF proposed by \citet{Izotov:06}, reaching up to $\sim$0.7 dex of difference. In contrast, the ICF by \citet{Rodriguez:05} seems to provide more suitable predictions, especially when considering $t^2 > 0$, where the differences are reduced to $\sim$0.2 dex. To resolve these ICF-discrepancies, an observational sample dedicated to the detection of \feiv~emission lines is necessary, which is extremely scarce in the literature.

Our results confirm the existence of dust trapping Fe within photoionized environments, in consistency with previous studies \citep{Rodriguez:02, Rodriguez:05, Izotov:06}. We estimate that at solar metallicities, $\sim$95\% of the Fe is trapped in dust grains, while at typical metallicities of the LMC, this factor is $\sim$75\%, and for those of the SMC, it is $\sim$35\%. If we consider that O may be trapped in solid compounds along with Fe in the ionized gas, this implies that there is up to $\sim$0.1 dex of O trapped in dust at solar metallicity, while in the LMC, this value decreases to $\sim$0.07 dex, and in the SMC, it decreases to $\sim$0.03 dex. At metallicities lower than those of the SMC, the fraction of O trapped in dust within \hii~regions could be negligible considering typical uncertainties in the determination of nebular chemical abundances. These values are upper limits and should be used with caution. Our results are in disagreement with the common assumption of adding 0.1 dex --regardless of the metallicity range or degree of ionization-- to nebular O/H determinations to account for the contribution of O trapped in dust grains.

Our distribution of gas-phase Fe/O abundance ratios in star-forming regions follows a moderate linear correlation with O/H abundance. In all cases, we have obtained subsolar values of Fe/O. The observed dispersion does not appear to be solely linked to errors in the \feiii~$\lambda 4658$ flux, but to other physical phenomena related to the creation and destruction of dust. When 12+log(O/H) $<$ 7.6, the Fe/O values seem to reach a plateau around log(Fe/O) $\approx-1.74$. If this plateau is confirmed by subsequent observations, it indicates that the fraction of dust trapping Fe is negligible at such low metallicities. This could help to explain, at least in part, why these lines have been detected in high-z galaxies revealed by the JWST \citep{ArellanoCordova:22}. The value to which regions with 12+log(O/H) $<$ 7.6 seem to converge is consistent with the observed Fe/O values in metal-poor stars in our Galaxy. This would be consistent with a high star formation efficiencies in these regions, which are dwarf irregular galaxies.

We report a stronger linear correlation between Fe/N and N/H compared to the analogous case involving O. We interpret this finding as indicative that the production timescales of N and Fe are similar, while dust formation trapping Fe scales with its own availability. The hardness of the ionizing radiation also plays an important role destroying dust grains and scales with the Fe abundance. The first hypothesis is consistent with the stellar abundances of Fe/N vs. N/H in the Galaxy, which follow a rather flat pattern in the abundance range covered in this study. Considering this hypothesis, we have derived Eq.~\eqref{eq:Fe_Dustratio_derived_N}, which allows us to approximately estimate the fraction of Fe trapped into dust. Using this equation along with the gas-phase Fe/H abundances, we have been able to  produce the relation between the total ISM (gas+dust) Fe/O ratio and O/H (see Fig.~\ref{fig:Total_result}). The use of N as an element that connects stellar and nebular metallicities is certainly a promising idea that requires an expansion of both stellar and nebular simultaneous determinations of Fe, O, and N, especially at low metallicities. On the other hand, directly comparing stellar Fe-metallicities and nebular O-metallicities using the solar reference, regardless of the abundance ranges studied, certainly is inadvisable.

While J0811+4730 and J1631+4426 do show slightly elevated Fe/O abundance ratios, these are consistent with other determinations in regions with similar O/H abundances within the errors and are certainly subsolar. The differences found are of the order of typical uncertainties associated with the Fe-ICF discussed here. Therefore, we do not observe clear evidence supporting the Fe enrichment produced by very massive stars ($M_\text{init} > 300M_{\odot}$). However, further observations and a better refinement of the Fe-ICF are necessary to reach solid conclusions on this regard. Finally, our analysis of J1205+4551 \citep{Izotov:17b, Izotov:21b} and the Sunburst Arc \citep{Welch:24}, the latter galaxy observed with the JWST at $z=2.37$, suggests that systems with high N/O ratios may also have relatively elevated Fe/O values. Since J1205+4551 exhibits evidence of the presence of WR stars, one would have to ask whether the high values of Fe/O and N/O in these regions may be linked to the WR activity.

\begin{acknowledgements}

We dedicate this work to the memory of our esteemed colleague and friend Dr. Claudio Mendoza Guardia (1951-2024). Among his many contributions to astronomy, he recently reviewed and improved the radiative and collisional atomic coefficients of Fe, which now allow us to conduct this study with the required precision. The authors thank the anonymous referee for their careful review and valuable comments. These have substantially helped to improve this article. JEMD and KK gratefully acknowledge funding from the Deutsche Forschungsgemeinschaft (DFG, German Research Foundation) in the form of an Emmy Noether Research Group (grant number KR4598/2-1, PI Kreckel) and the European Research Council’s starting grant ERC StG-101077573 (“ISM-METALS").
AACS is supported by the German \emph{Deut\-sche For\-schungs\-ge\-mein\-schaft, DFG\/} in the form of an Emmy Noether Research Group -- Project-ID 445674056 (SA4064/1-1, PI Sander). AACS further acknowledges funding provided by the Federal Ministry of Education and Research (BMBF) and the Baden-Württemberg Ministry of Science as part of the Excellence Strategy of the German Federal and State Governments.
CE and JG-R acknowledge support from the Agencia Estatal de Investigaci\'on del Ministerio de Ciencia e Innovaci\'on (AEI- MCINN) under grant Espectroscop\'ia de campo integral de regiones H II locales. Modelos para el estudio de regiones H II extragala\'acticas with reference 10.13039/501100011033 and from  grant P/308614 financed by funds transferred from the Spanish Ministry of Science, Innovation and Universities, charged to the General State Budgets and with funds transferred from the General Budgets of the Autonomous Community of the Canary Islands by the MCIU. JG-R also acknowledges funds from the Spanish Ministry of Science and Innovation (MICINN) through the Spanish State Research Agency, under Severo Ochoa Centres of Excellence Program 2020-2023 (CEX2019-000920-S). IDL and SvdG acknowledge funding from the European Research Council (ERC) under the European Union's Horizon 2020 research and innovation program DustOrigin (ERC-2019-StG-851622) and from the Flemish Fund for Scientific Research (FWO-Vlaanderen) through the research project G023821N. SFS thanks the PAPIIT-DGAPA AG100622 project and CONACYT grant CF19-39578. 

\end{acknowledgements}


\bibliographystyle{aa}
\bibliography{refs} % if your bibtex file is called example.bib

\begin{appendix} %First appendix

\section{The empirical relationship between metallicity and reddening in star-forming nebulae.}
\label{sec:appendix_CB}

\begin{figure}[h]
\centering    
\includegraphics[width=\hsize ]{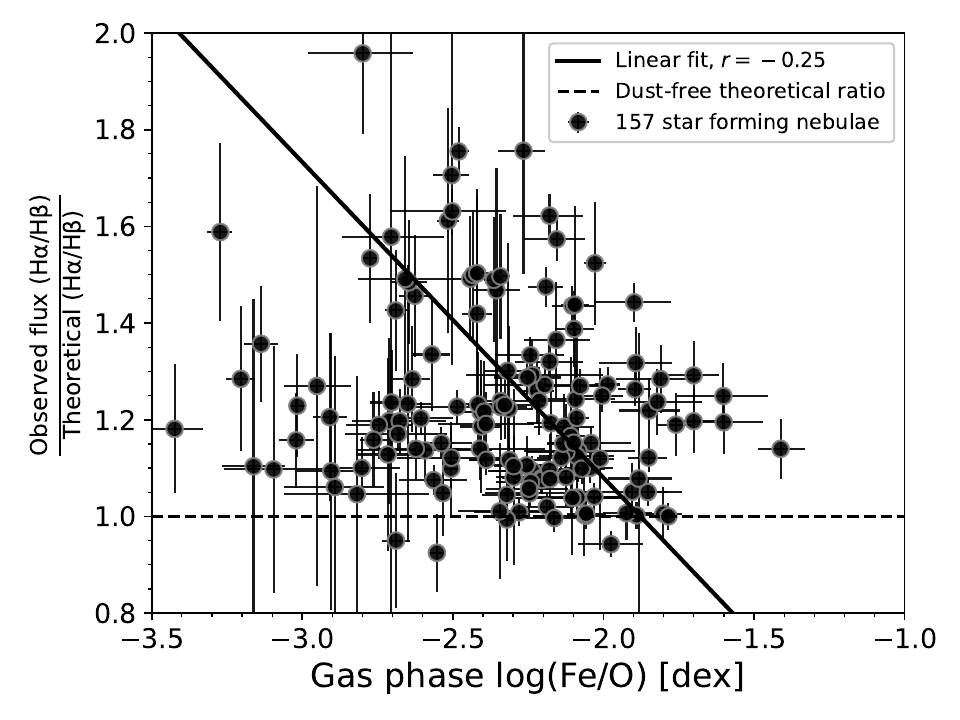}
\caption{Comparison of ratios between the observed and theoretical H$\alpha$/H$\beta$ fluxes with the gas-phase Fe/O values analyzed in this manuscript for a subsample with reported flux values prior to reddening correction.} 
\label{fig:redd_FEO}
\end{figure}

\begin{figure}[h]
\centering    
\includegraphics[width=\hsize ]{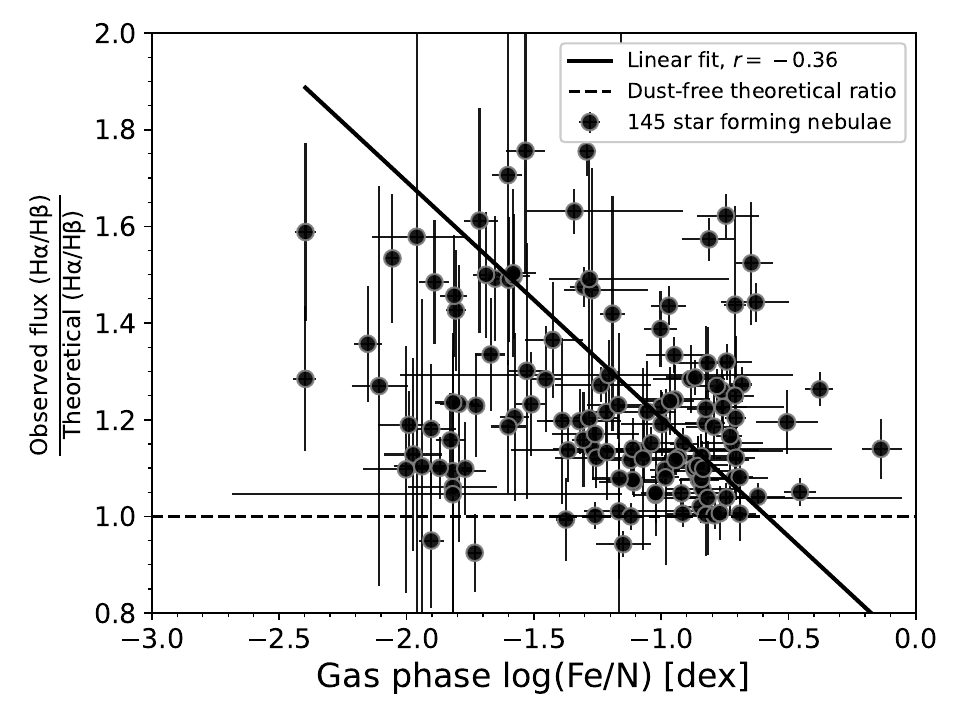}
\caption{Same as Fig.~\ref{fig:redd_FEO} but for the gas-phase Fe/N abundances.}
\label{fig:redd_FEN}
\end{figure}

Typically, the reddening corrections in star-forming regions are traced by comparing the observed H$\alpha$/H$\beta$ fluxes with the theoretically expected values, using an extinction law suited for each specific region \citep{Calzetti:94}. In a subsample of our star-forming regions, we have access to the observed fluxes before reddening correction. In such regions, we compared the observed H$\alpha$/H$\beta$ values with the theoretical predictions of \citet{Storey:95}, using PyNeb \citep{Luridiana:15}. We considered the density and temperature $T_e$(\oiii) (the impact of $t^2$ is negligible in this RL-ratio) adopted for each region, presented in Tables \ref{table:densities} and \ref{table:temperatures_adopted}, respectively. Then, we compared the ratios between the observed and theoretical H$\alpha$/H$\beta$ fluxes with the gas-phase Fe/O and Fe/N values analyzed in this manuscript, as shown in Figs. \ref{fig:redd_FEO} and \ref{fig:redd_FEN}. The numerical values are presented in Table~\ref{table:Ha_Hb}

Unfortunately, not all the reference spectra in Table~\ref{table:IDs} explicitly report the observed flux values before reddening correction, and the statistics presented in Figs. \ref{fig:redd_FEO} and \ref{fig:redd_FEN} are notably less robust than those shown in Figs. \ref{fig:resulting_FeO} and \ref{fig:resulting_FeN}. However, there is a weak anticorrelation between Fe/N and the H$\alpha$/H$\beta$ flux ratios, as specifically shown in Fig.~\ref{fig:redd_FEN}. This anticorrelation is expected if there is a connection between the dust present within the \hii~regions and the dust in the surrounding neutral gas, which should cause most of the optical extinction. In the global trend, when log(Fe/N) $\sim$ -0.5 dex, H$\alpha$/H$\beta$ appears to reach values consistent with the theoretical fluxes under conditions of no extinction. These Fe/N values are achieved in low-metallicity regions, near the limit of 12+log(N/H) $\sim$ 6.3 (see Table~\ref{table:equations_Fe_O_N}). A qualitatively similar situation occurs with Fe/O, although the correlation with H$\alpha$/H$\beta$ is evidently less strong than in the case of Fe/N. This supports the interpretation made in the present article regarding the distribution of gas-phase Fe/N and Fe/O abundances in terms of dust depletion and the apparent closer correlation between Fe and N compared to Fe and O. Observed flux (H$\alpha$/H$\beta$) / Theoretical (H$\alpha$/H$\beta$) is an observational quantity independent of the specific extinction law used to model the reddening coefficient, c(H$\beta$), or the color excess $E(B-V)$. However, this can, in turn, increase the observed dispersion.

\begin{figure}[h]
\centering    
\includegraphics[width=\hsize ]{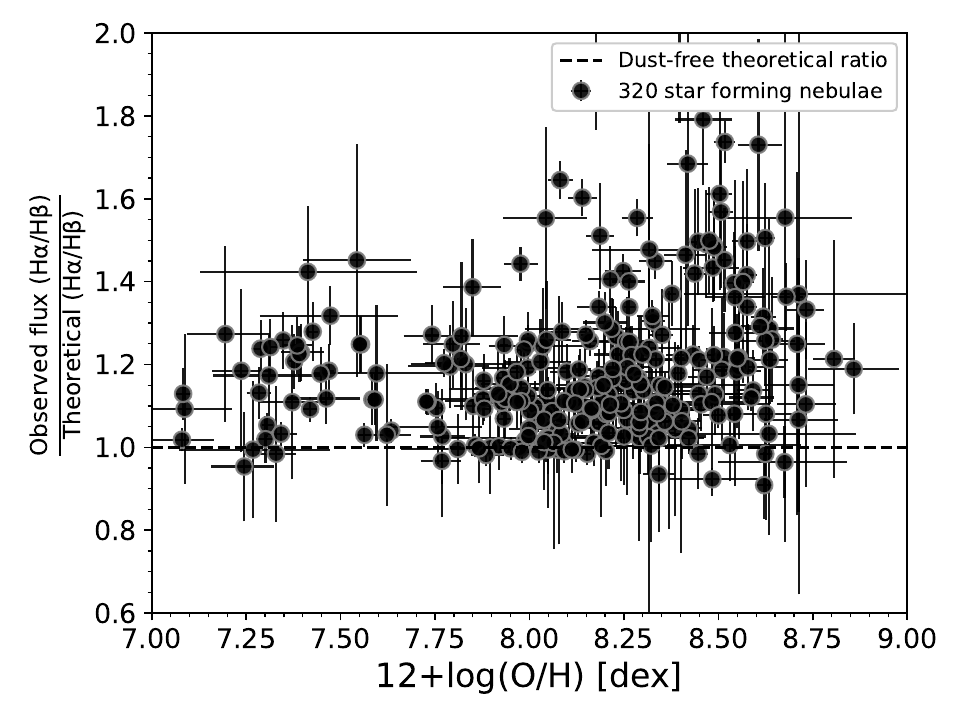}
\caption{Comparison of ratios between the observed and theoretical H$\alpha$/H$\beta$ fluxes with the 12+log(O/H) values for a subsample of the DESIRED database with reported flux values prior to reddening corrections.} 
\label{fig:redd}
\end{figure}

As a final note of interest, in the general DESIRED sample, we found that statistically, there are higher extinctions in regions with higher O/H abundance, as shown in Fig.~\ref{fig:redd}. The sample presented in that figure is larger than the one shown in Figs. \ref{fig:redd_FEO} and \ref{fig:redd_FEN}, as it includes regions without detections of \feiii$\lambda 4658$. It is notable that although the effect of extinction is certainly lower in regions of lower metallicity, it is still not negligible in most cases. Considering the discussion presented in Sec.~\ref{sec:discussion}, this could indicate that \hii~regions have a lower fraction of Fe trapped in dust than the surrounding neutral clouds. This could support the idea that part of the dust initially present in neutral gas clouds is destroyed during the ionization stage when massive stars form in \hii~regions. However, a systematic and simultaneous multi-wavelength study of Fe depletions in neutral and ionized gas in a statistically robust sample is necessary to reach more solid conclusions.

\FloatBarrier

\section{The impact of extrapolating $T_e$-$T_e$ relations in the Fe/O and Fe/N distributions}
\label{sec:appendix_a0}

Due to their ionization potential, Fe$^{2+}$, O$^{+}$, and N$^{+}$ are considered low ionization degree ions. This implies that determining their ionic abundances requires adopting a characteristic temperature for their coexistence volume. Various studies indicate that one of the most reliable diagnostics for this parameter is $T_e$(\nii~$\lambda 5755/\lambda 6584$) due to its low dependence on $n_e$ \citep{Esteban:09, ArellanoCordova:2020b, MendezDelgado:23b, RickardsVaught:24}. However, observationally, the detection of \nii~$\lambda 5755$ is particularly challenging for low metallicity regions, which often exhibit high ionization states \citep{Vilchez:88a}. Because of this, in most cases, a temperature relation is required to connect other temperature diagnostics with $T_e$(\nii~$\lambda 5755/\lambda 6584$). As described in Sec.~\ref{sec:obs}, this work adopted the relations proposed by \citet{Garnett:92} and \citet{MendezDelgado:23a} to connect $T_e$(\oiii~$\lambda 4363/\lambda 5007$) and/or $T_e$(\siii~$\lambda 6312/\lambda 9069$) when $T_e$(\nii~$\lambda 5755/\lambda 6584$) was not measured directly both in the case $t^2=0$ and $t^2>0$, respectively.

The use of these temperature relations is a potential source of systematic errors if they have significant deviations from what is present in real nebulae. Although this has been studied in several articles \citep{Skillman:03, Croxall:15, Berg:20, Rogers:21}, both in the case of $t^2=0$ \citep{Garnett:92} and in the case of $t^2>0$ \citep{MendezDelgado:23a}, there is a very limited number of low-metallicity regions with simultaneous determinations of $T_e$(\nii~$\lambda 5755/\lambda 6584$) and $T_e$(\oiii~$\lambda 4363/\lambda 5007$) or $T_e$(\siii~$\lambda 6312/\lambda 9069$). These limitations will be alleviated in upcoming studies (Orte-García M. et al., in prep) using the DESIRED sample, but the calibrations still lack statistical robustness for temperatures above 13,000K.

To quantify the impact that a hypothetical systematic deviation between the assumed temperature relations and those actually present in the nebulae could have on our determinations of Fe/N and Fe/O abundances, we can consider the following relation \citep{Stasinska:23}:

\begin{equation}
\label{eq:emissivity_and_intensity}
    \frac{\text{Fe}^{2+}}{\text{O}^+} = \frac{\left( \frac{I(\text{\feiii } \lambda 4658)}{I(\text{\oii }\lambda \lambda 3727 + 3729)} \right)}{\left( \frac{j(\text{\feiii } \lambda 4658)}{j(\text{\oii } \lambda \lambda 3727 + 3729)} \right)},
\end{equation}

where $I(\lambda)$ represents the observed line intensity and $j(\lambda)$ is the emissivity of that line. In most cases, \oii~$\lambda \lambda 3727+3729$ is the basis for estimating the abundances of O$^{+}$/H$^{+}$, and \feiii~$\lambda 4658$ is the basis for estimating the abundance of Fe$^{2+}$/H$^{+}$ in all cases. If $T_e$(\nii) presents a systematic error, it will affect both $j(\text{\oii } \lambda \lambda 3727 + 3729)$ and $j(\text{\feiii } \lambda 4658)$. We can quantify that systematic error using PyNeb \citep{Luridiana:15} and Eq.~\eqref{eq:emissivity_and_intensity} as follows:

\begin{equation}
\label{eq:bias_test}
\frac{(\text{Fe}^{2+}/\text{O}^+)_{\text{Biased-$T_e$}}}{(\text{Fe}^{2+}/\text{O}^+)_{\text{True-$T_e$}}} = \frac{\left( \frac{j(\text{\feiii } \lambda4658)}{j(\text{\oii } \lambda \lambda 3727+3729)} \right)_{\text{True-$T_e$}}}{\left( \frac{j(\text{\feiii } \lambda4658)}{j(\text{\oii } \lambda\lambda3727+3729)} \right)_{\text{Biased-$T_e$}}}.
\end{equation}

We will consider the following case: In low metallicity regions, only $T_e$(\oiii) is available and we must use the relationships from \citet{Garnett:92} and \citet{MendezDelgado:23a} to determine $T_e$(\nii) and $T_0$(O$^{2+}$). We also assume that the $T_e$ relationships induce a systematic error of 20\%. This value is notably higher than the uncertainties in the slope and intercept of the $T_e$(\oiii) - $T_e$(\nii) relations derived by \citet{MendezDelgado:23b} and $T_e$(\nii) - $T_0$ (O$^{2+}$) derived by \citet{MendezDelgado:23a}, which are around $\sim$5\% in the slopes and $\sim$10\% in the intercepts.

\begin{figure}[h]
\centering    
\includegraphics[width=\hsize ]{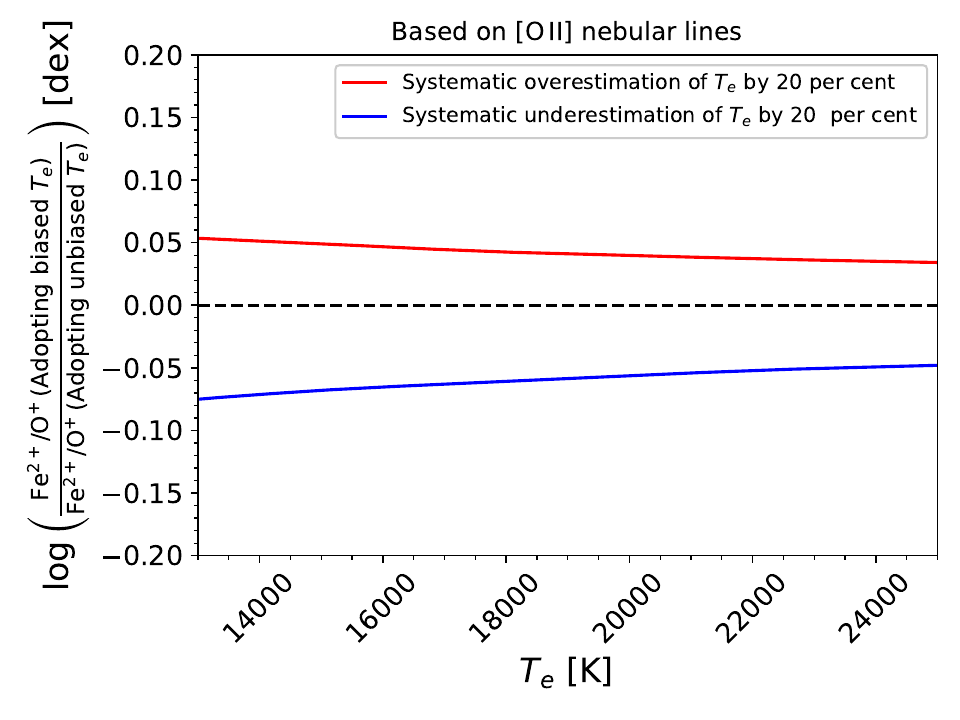}
\caption{Upper limits to the systematic errors in the abundance of Fe$^{2+}$/O$^{+}$ considering a hypothetical systematic error in $T_e$ of 20\%. These limits have been calculated using PyNeb \citep{Luridiana:15} and Eq.~\eqref{eq:bias_test}.} 
\label{fig:Bias_Fe3O2}
\end{figure}

\begin{figure}[h]
\centering    
\includegraphics[width=\hsize ]{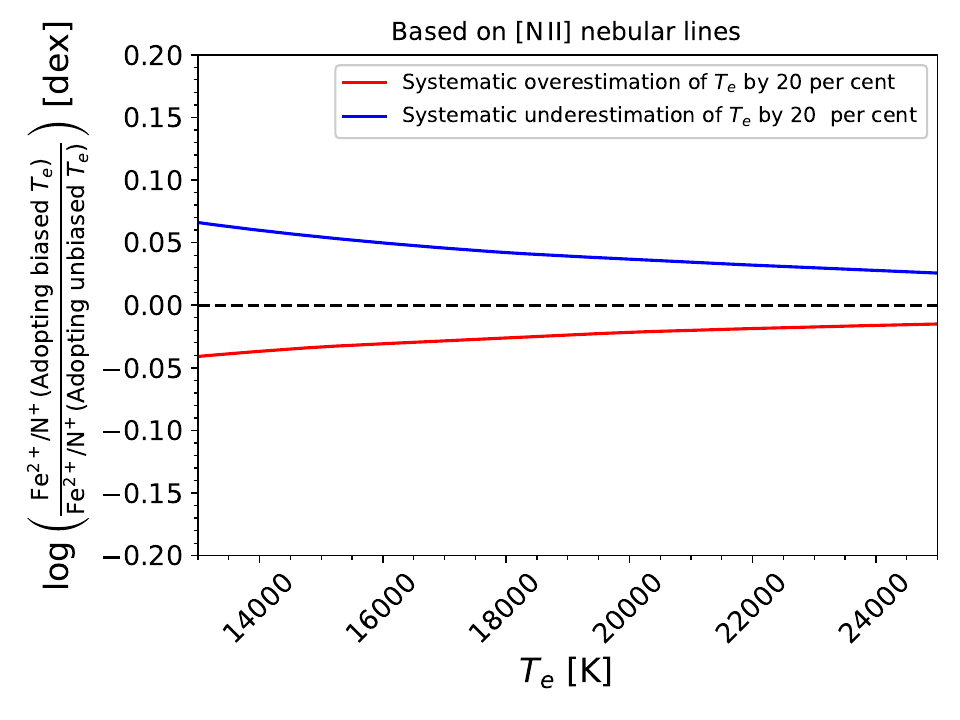}
\caption{Same as Fig.~\ref{fig:Bias_Fe3O2} but considering the Fe$^{2+}$/N$^{+}$ abundances.} 
\label{fig:Bias_Fe3N2}
\end{figure}

In Fig.~\ref{fig:Bias_Fe3O2} it is shown that the potential impact of a systematic error in $T_e$ of 20\% on the ionic abundance of Fe$^{2+}$/O$^{+}$ is $\sim$0.05 dex when adopting the nebular lines \oii$\lambda \lambda 3727+3729$. In this figure, $n_e=100$ cm$^{-3}$ was considered, although the conclusions do not depend on the value of this parameter. Despite considering systematic errors of several thousand degrees Kelvin, the impact on Fe$^{2+}$/O$^{+}$ is very small due to the similarity between the excitation energies of the $^3\text{F}_4$ atomic level of Fe$^{2+}$ ($\sim 2.7$ eV) and the $^2\text{D}$ levels of O$^{+}$ ($\sim 3.3$ eV). The same is true for the abundance of Fe$^{2+}$/N$^{+}$ (we used \nii~6584, which arises from the $^1\text{D}$ level of N$^{+}$ ($\sim 1.9$ eV)) as shown in Fig.~\ref{fig:Bias_Fe3N2}.

\begin{figure}[h]
\centering    
\includegraphics[width=\hsize ]{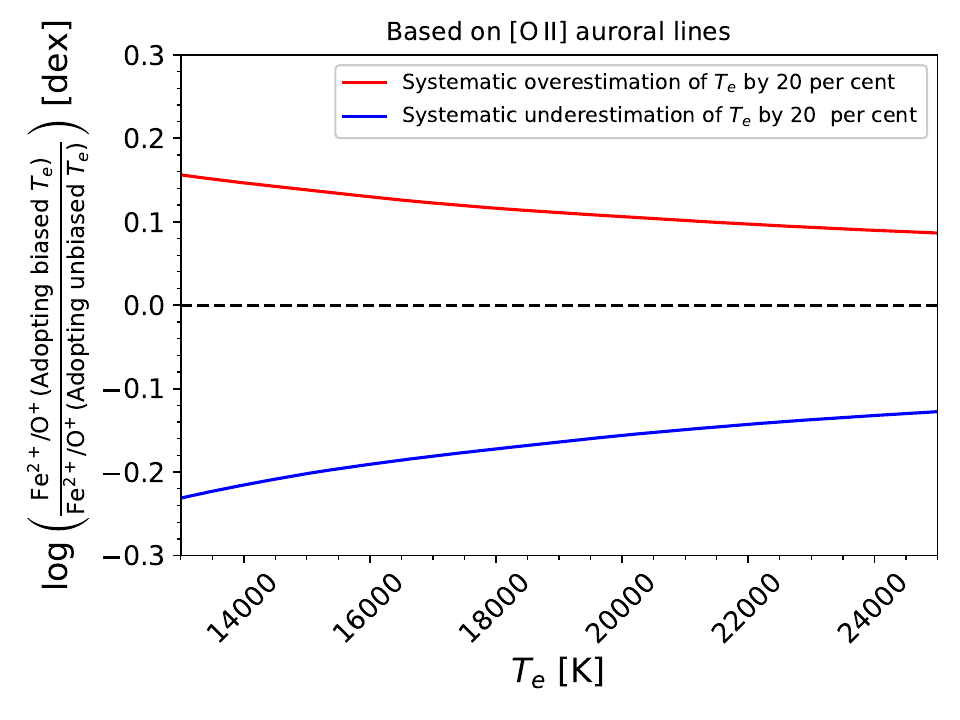}
\caption{Same as Fig.~\ref{fig:Bias_Fe3O2} but considering the determination of the O$^{+}$/H$^{+}$ ionic abundances based on the \oii~$\lambda \lambda 7319+7320+7330+7331$ auroral lines.} 
\label{fig:Bias_Fe3O2_auroral}
\end{figure}

\begin{figure}[h]
\centering    
\includegraphics[width=\hsize ]{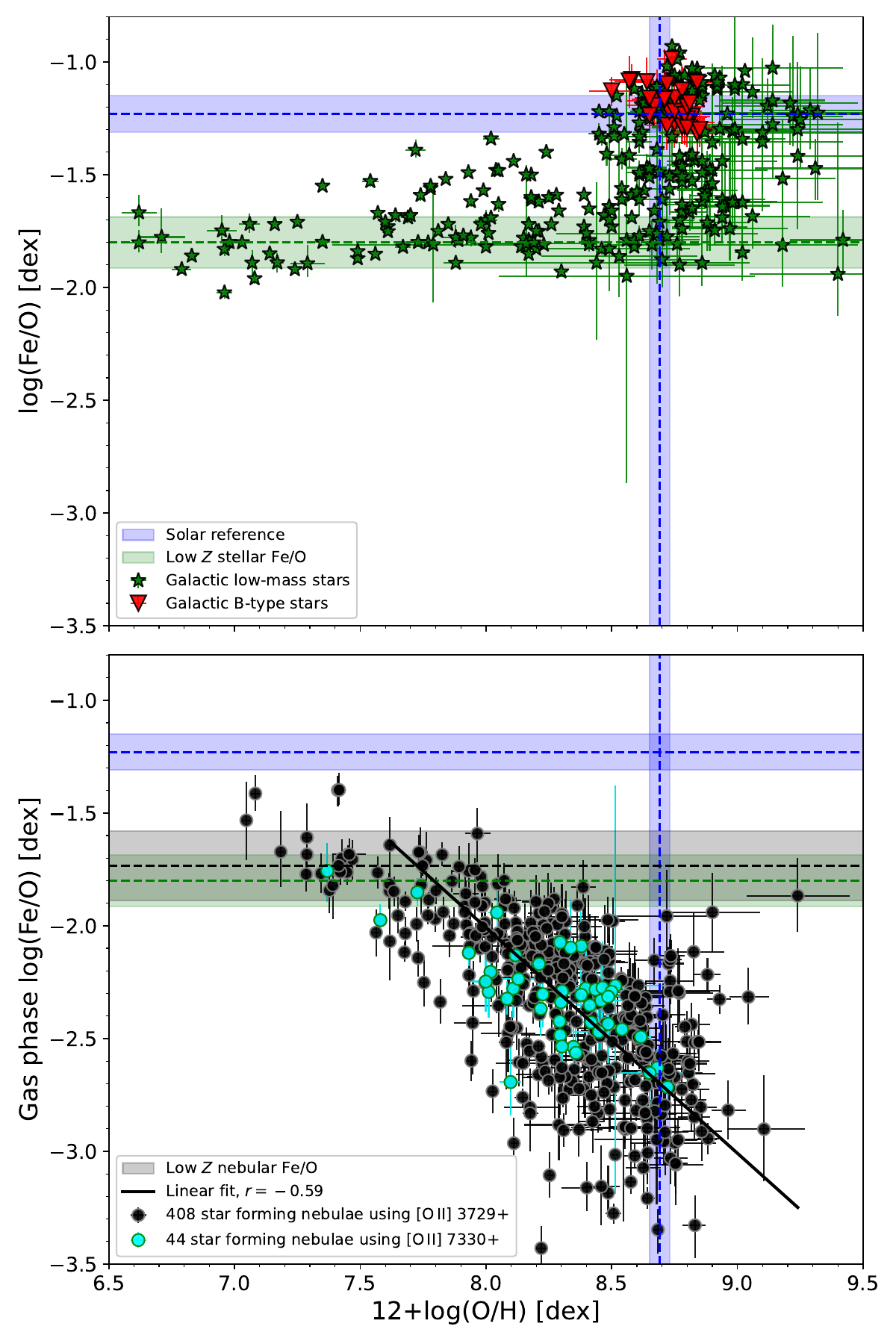}
\caption{Same as Fig.~\ref{fig:resulting_FeO} but highlighting the subset of objects where the determination of the O$^+$ abundance relies on the \oii~auroral lines.} 
\label{fig:resulting_FeO_highlight_auroral}
\end{figure}

As shown in Fig.~\ref{fig:Bias_Fe3O2_auroral}, the situation is notably different when the abundance of O$^{+}$ has been determined using the auroral lines of \oii~$\lambda \lambda 7319+7320+7330+7331$. The temperature dependence becomes very significant, potentially introducing systematic errors greater than $\sim$0.1 dex. However, only 44 regions (<10\% of the entire sample) have O$^{+}$ determinations based on these auroral lines of \oii. In Fig.~\ref{fig:resulting_FeO_highlight_auroral} we demonstrate that these regions do not exhibit anomalies in the observed trends and that our conclusions do not rely particularly on these regions.

\begin{figure}[h]
\centering    
\includegraphics[width=\hsize ]{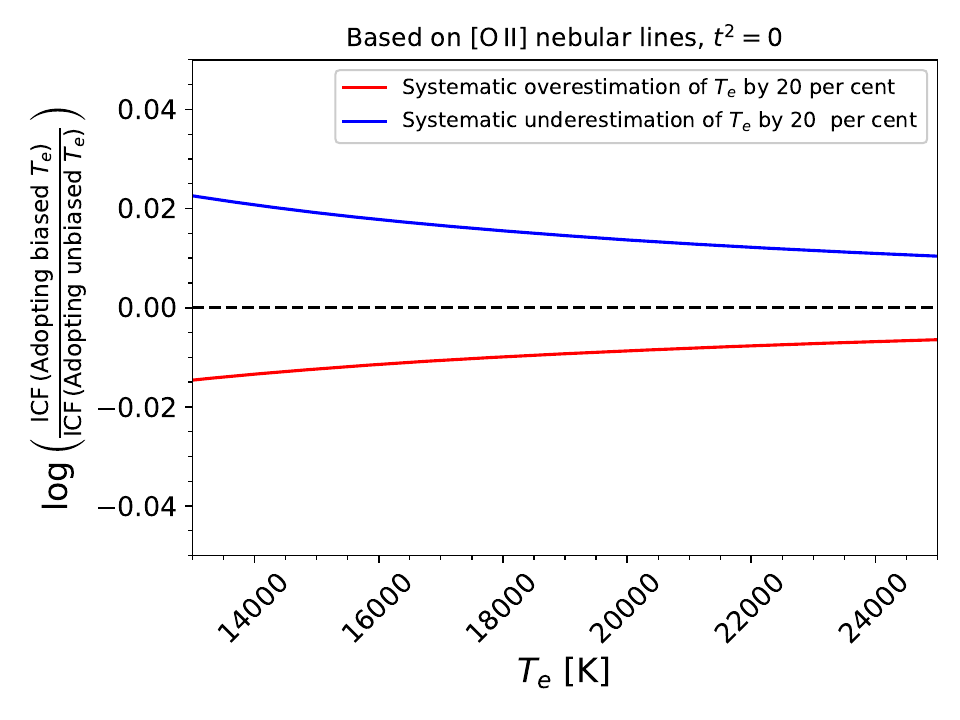}
\caption{Upper limits to the systematic errors in ICF(Fe) from \citet{Rodriguez:05} considering a hypothetical systematic error in $T_e$ of 20\%. These limits have been calculated using PyNeb \citep{Luridiana:15} and Eq.~\eqref{eq:bias_test}.} 
\label{fig:bias_ICF_FeO}
\end{figure}

\begin{figure}[h]
\centering    
\includegraphics[width=\hsize ]{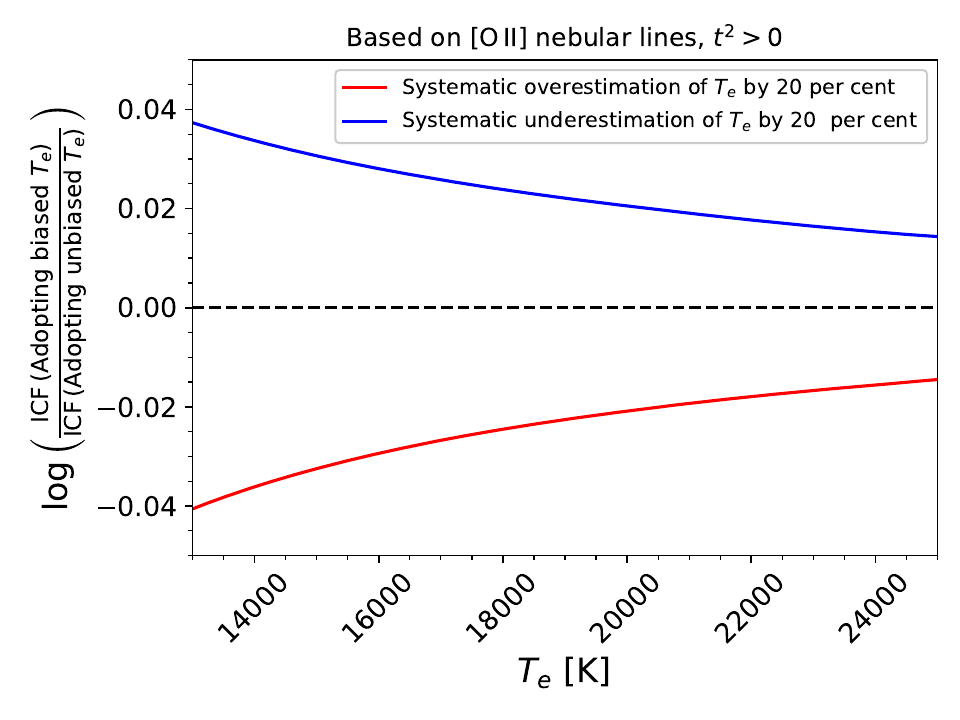}
\caption{Same as Fig.~\ref{fig:bias_ICF_FeO} but considering the case $t^2>0$.} 
\label{fig:bias_ICF_FeO_t2}
\end{figure}

To explore the bias in the total abundances of Fe/O and Fe/N, it is necessary to consider the ICF and its relationship with the ionic abundances of O$^+$ and O$^{2+}$, both in the case of $t^2=0$ and $t^2>0$. In the case of ICF(Fe), we consider the ICF by \citet{Rodriguez:05}. As shown in Figs.~\ref{fig:bias_ICF_FeO} and \ref{fig:bias_ICF_FeO_t2}, the potential systematic error is smaller than $\sim$0.04 dex and acts in the opposite direction to the systematic error in the Fe$^{2+}$/O$^{+}$ abundance presented in Fig.~\ref{fig:Bias_Fe3O2}, which causes the systematic errors to partially cancel each other when determining Fe/O = ICF $\times$ Fe$^{2+}$/O$^{+}$.

\begin{figure}[h]
\centering    
\includegraphics[width=\hsize ]{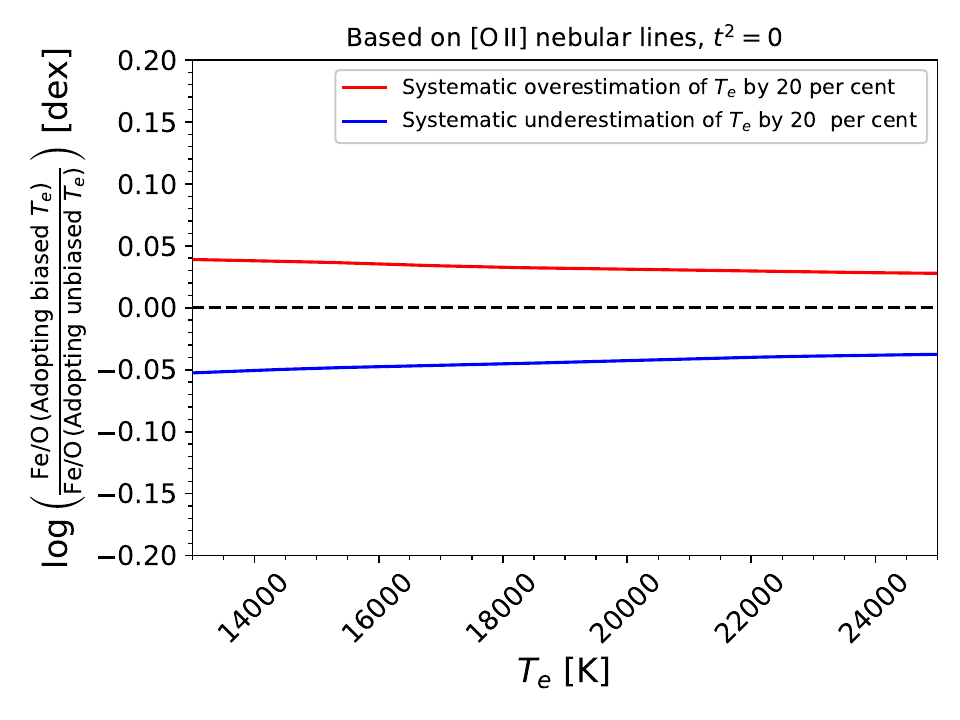}
\caption{Same as Fig.~\ref{fig:Bias_Fe3O2} but considering Fe/O=ICF $\times$ Fe$^{2+}$/O$^{+}$ and the propagation of their corresponding systematic errors under the case $t^2=0$.} 
\label{fig:total_bias_FEO}
\end{figure}

\begin{figure}[h]
\centering    
\includegraphics[width=\hsize ]{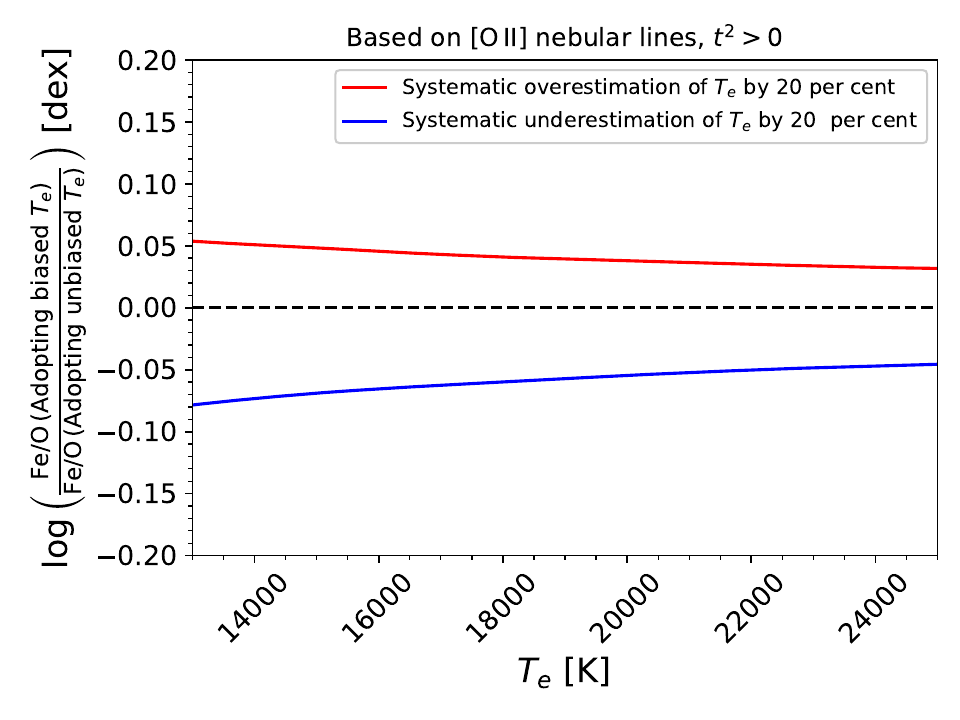}
\caption{Same as Fig.~\ref{fig:total_bias_FEO} but considering the case $t^2>0$.} 
\label{fig:total_bias_t2_FEO}
\end{figure}

Considering Figures \ref{fig:Bias_Fe3O2}, \ref{fig:bias_ICF_FeO} and \ref{fig:bias_ICF_FeO_t2}, we can estimate the total impact of a hypothetical bias in $T_e$ on the Fe/O abundances. This is shown in Figures \ref{fig:total_bias_FEO} and \ref{fig:total_bias_t2_FEO}. These figures demonstrate that the potential systematic error induced in the Fe/O abundances due to very large hypothetical errors in $T_e$ from the extrapolation of the temperature relationships by \citet{Garnett:92} and \citet{MendezDelgado:23a} in low-metallicity regions is around $\sim$0.05 dex. This rules out the possibility that the apparent plateau observed in the Fe/O distribution at low metallicities (see Fig.~\ref{fig:resulting_FeO}) is the result of a systematic error in the inferred $T_e$(\nii).

\begin{figure}[h]
\centering    
\includegraphics[width=\hsize ]{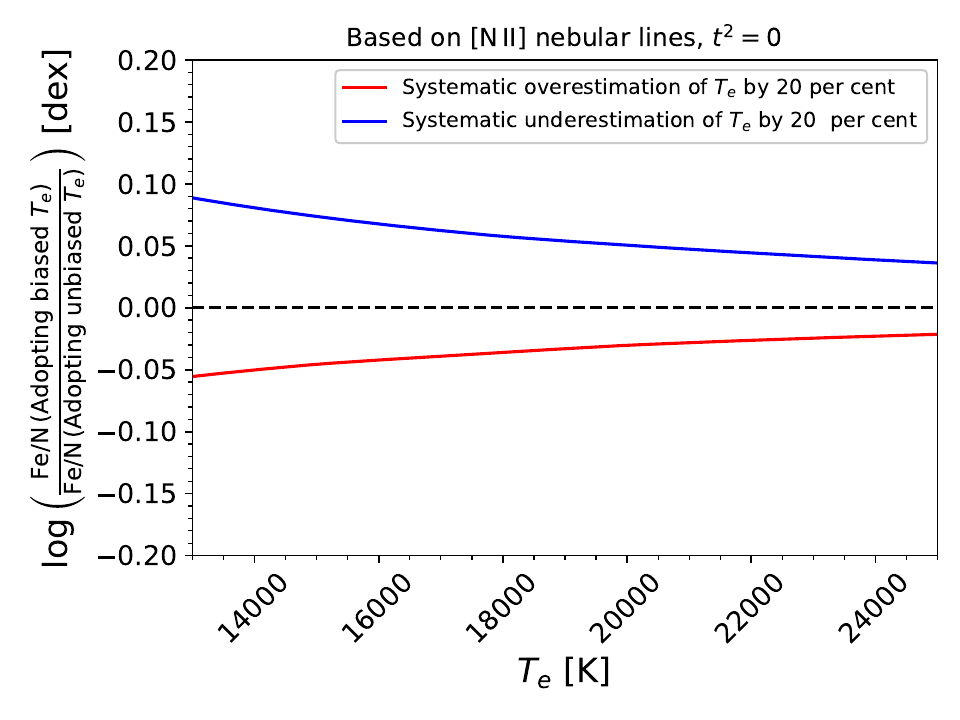}
\caption{Same as Fig.~\ref{fig:total_bias_FEO} but considering Fe/N under the case $t^2=0$.} 
\label{fig:total_bias_FEN}
\end{figure}

\begin{figure}[h]
\centering    
\includegraphics[width=\hsize ]{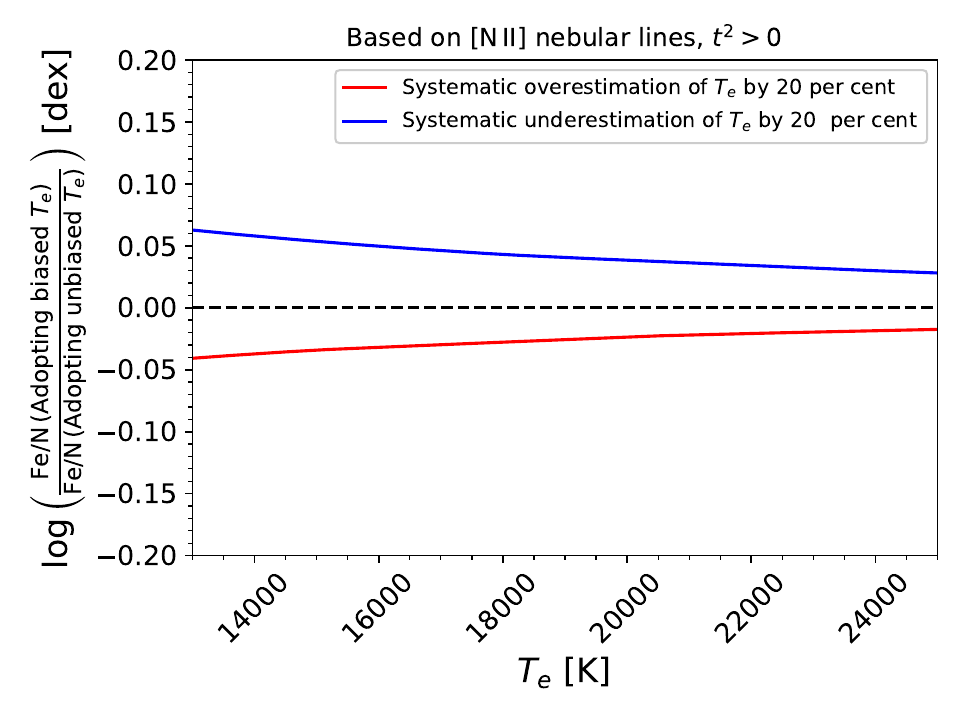}
\caption{Same as Fig.~\ref{fig:total_bias_FEN} but considering the case $t^2>0$.} 
\label{fig:total_bias_t2_FEN}
\end{figure}

The case of the total Fe/N abundance involves two ICFs: the one from \citet{Rodriguez:05} for Fe and the one from \citet{Amayo:21} for N. In the latter case, the equation is a fifth-degree polynomial over $\omega=\text{O}^{2+}/(\text{O}^{+}+\text{O}^{2+})$. For low metallicity regions, $\omega$ is expected to be close to 1 \citep{Vilchez:88a}, minimizing any potential bias in the ionic abundances induced by $T_e$-errors. As a good approximation, we can assume that the ICF(N) induce a negligible systematic error due to systematics on $T_e$. Such case is presented in figures \ref{fig:total_bias_FEN} and \ref{fig:total_bias_t2_FEN} for the case $t^2=0$ and $t^2>0$, respectively. The potential systematic bias in the Fe/N distribution is rather similar than in the case of Fe/O, being up to $\sim$0.05 dex.

\FloatBarrier

\section{Fe/O and Fe/N distributions considering $t^2=0$ }
\label{sec:appendix_a}

The shape of the Fe/O and Fe/N distributions does not change in the case of $t^2=0$. In these cases, the observational fits to the star-forming nebulae are as follows:

- For 12+log(O/H) < 7.6:
\[
\text{log(Fe/O)} = -0.75 \pm 0.19
\]

- For 12+log(O/H) $\geq$ 7.6:
\[
\text{log(Fe/O)} = (-1.02 \pm 0.05) \times [12+\text{log(O/H)}] + (5.97 \pm 0.39)
\]

- For 12+log(N/H) < 6.3:
\[
\text{log(Fe/N)} = -0.50 \pm 0.16
\]

- For 12+log(N/H) $\geq$ 6.3:
\[
\text{log(Fe/N)} = (-0.91 \pm 0.03) \times [12+\text{log(N/H)}] + (5.20 \pm 0.20)
\]

This results in Eq~\eqref{eq:Fe_Dustratio_derived_N_t2}.

\begin{equation}
\label{eq:Fe_Dustratio_derived_N_t2}
\frac{\text{Fe}_{\text{Dust}}}{\text{Fe}_{\text{Total}}}_{t^2=0} \approx 1-6\times 10^{-6} \times \left( \frac{\text{N}}{\text{H}}\right)^{-0.91} . 
\end{equation}

\begin{figure}[h]
\centering    
\includegraphics[width=\hsize ]{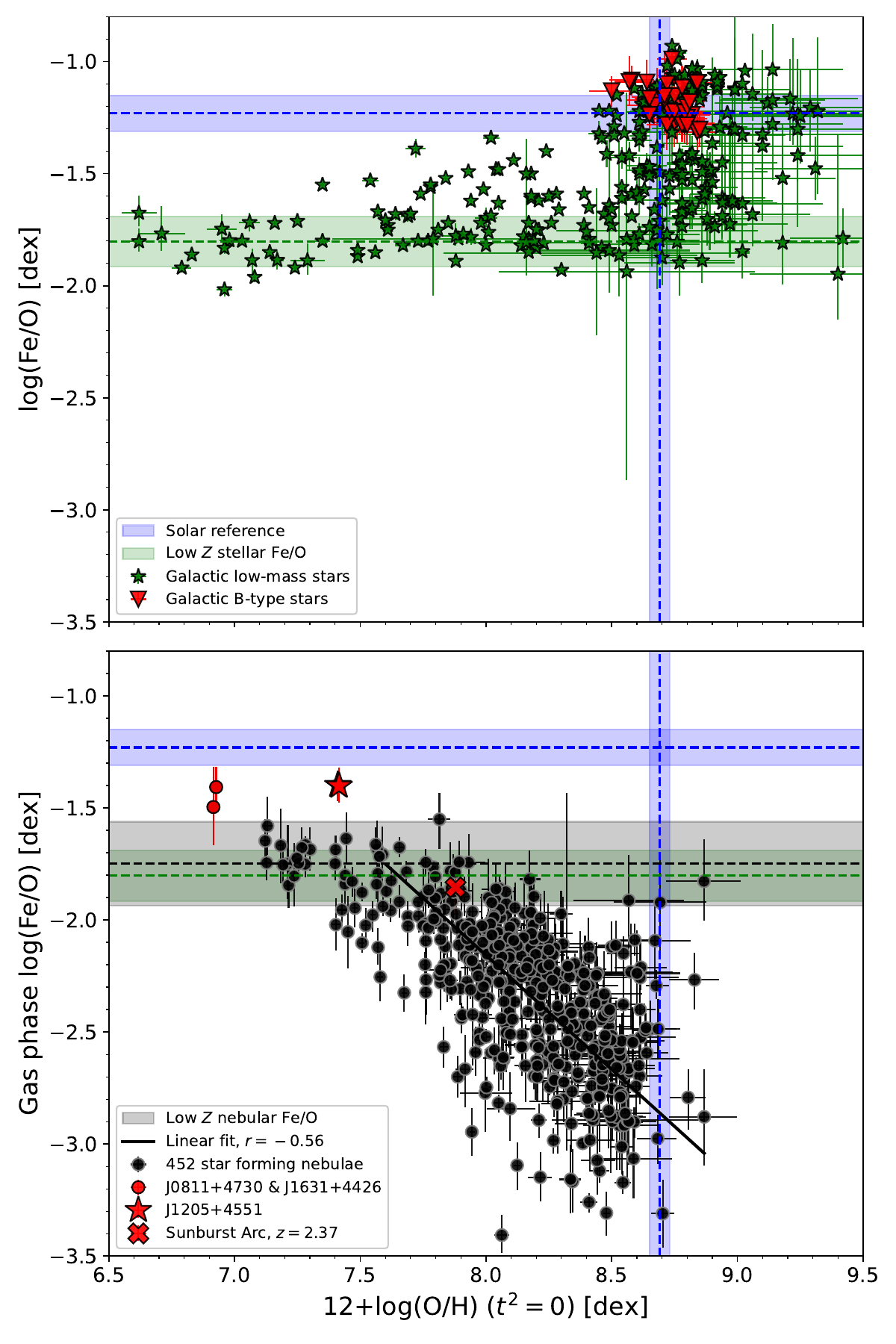}
\caption{Same as Fig.~\ref{fig:resulting_FeO} but considering an homogeneous nebular temperature $t^2=0$.} 
\label{fig:resulting_FeOApp}
\end{figure}

\begin{figure}[h]
\centering    
\includegraphics[width=\hsize ]{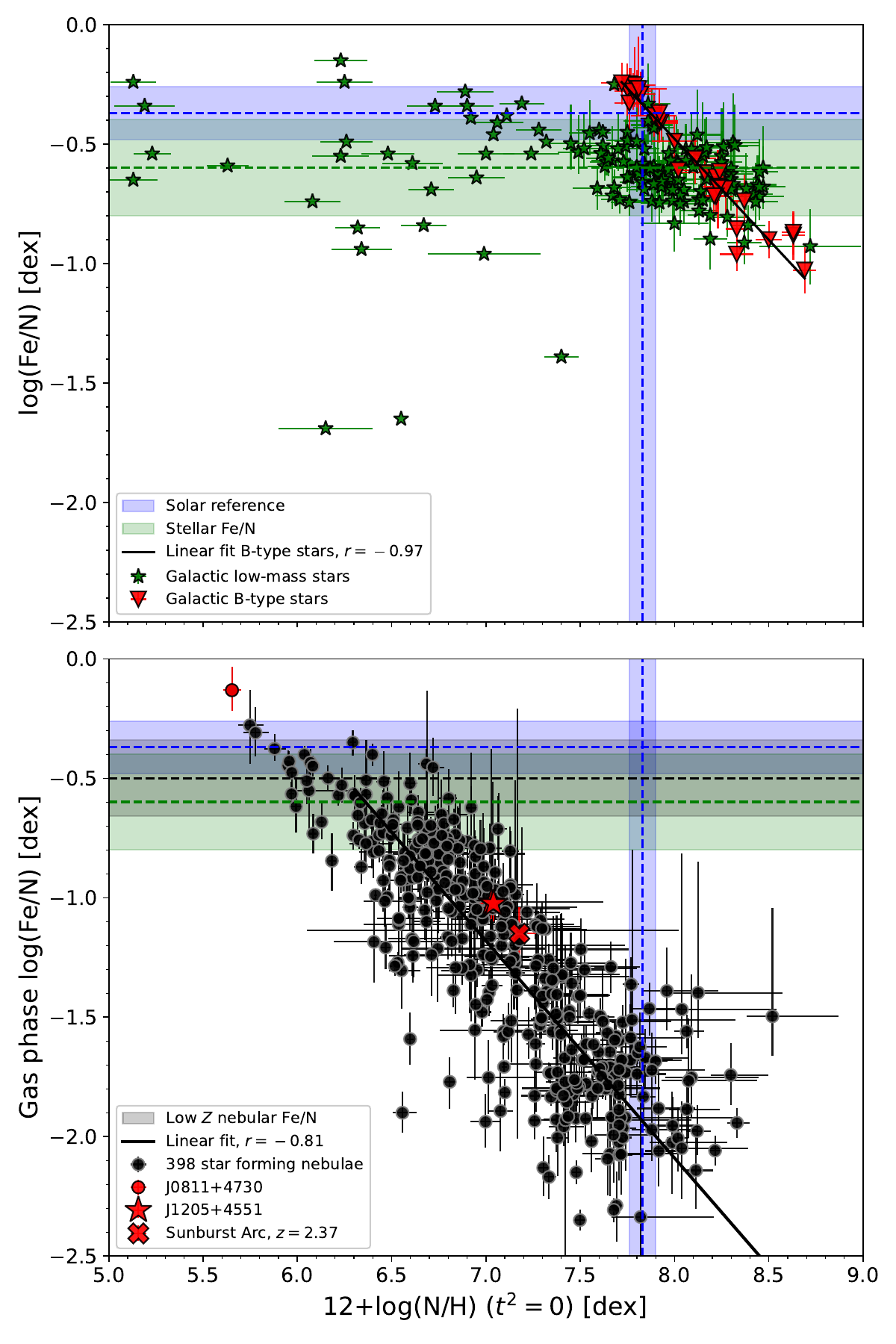}
\caption{Same as Fig.~\ref{fig:resulting_FeN} but considering an homogeneous nebular temperature $t^2=0$.} 
\label{fig:resulting_FeNApp}
\end{figure}

\FloatBarrier

\section{Tables of values and references}

\begin{table*}
    \caption{Atomic data set used for collisionally excited lines.}
    \label{table:atomic_data}
\begin{tabular}{lll}
        \hline
        \noalign{\smallskip}
        Ion & Transition probabilities & Collision strengths\\
        \noalign{\smallskip}
        \hline
        \noalign{\smallskip}
            O$^{+}$   &  \citet{FroeseFischer:04} & \citet{Kisielius:09}\\
            O$^{2+}$  &  \citet{Wiese:96}, \citet{Storey:00} & \citet{Storey:14}\\
            N$^{+}$   &  \citet{FroeseFischer:04} & \citet{Tayal:11}\\
            S$^{+}$   &  \citet{Irimia:05} & \citet{Tayal:10}\\
            S$^{2+}$  &  \citet{FroeseFischer:06} & \citet{Grieve:14}\\
            Cl$^{2+}$ &  \citet{Fritzsche:99} & \citet{Butler:89}\\
            Ar$^{2+}$ &   \citet{Mendoza:83b}, \citet{Kaufman:86}  & \citet{Galavis:95}\\
            Ar$^{3+}$ &   \citet{Mendoza:82b}  & \citet{Ramsbottom:97}\\
            Fe$^{+}$ & \citet{Bautista:15}, \citet{Mendoza:23}& \citet{Bautista:15}, \citet{Mendoza:23}\\
            Fe$^{2+}$ & \citet{Deb:09}, \citet{Mendoza:23} & \citet{Zhang:96}, \citet{Mendoza:23}\\
            Fe$^{3+}$ & \citet{FroeseFischer:08}& \citet{Zhang:97}\\
            Fe$^{4+}$ & \citet{Nahar:00}& \citet{Ballance:07}\\
        \noalign{\smallskip}
        \hline
    \end{tabular}
\end{table*}

%Gal. & Reg. & $n_e$ & $T_e$(\oiii) & $T_e$(\siii)   & $T_e$(\nii)  &  Reference \\
% &  &  [cm$^-3$] & [K] &  [K] & [K] &   \\

\longtab[2]{
\begin{longtable}{cccccc}
\caption{Reference number, galaxy, region name and references to the sample of nebular objects analyzed in this study.}\\
\hline
\hline
Reference number& Galaxy & Region &  Reference \\
\hline
\endfirsthead
\caption{Reference number, galaxy, region name and references to the sample of nebular objects analyzed in this study.}\\
\hline
Reference number & Galaxy & Region &  Reference \\
\hline
\endhead
\hline
\endfoot
\hline
\endlastfoot
\label{table:IDs}
D1 & 0275-51910-445 & - &\citet{Izotov:06}\\
D2 & 0308-51662-081 & - &\citet{Izotov:06}\\
D3 & 0329-52056-633 & - &\citet{Izotov:06}\\
D4 & 0330-52370-471 & - &\citet{Izotov:06}\\
D5 & 0337-51997-097 & - &\citet{Izotov:06}\\
D6 & 0358-51818-504 & - &\citet{Izotov:06}\\
D7 & 0364-52000-187 & - &\citet{Izotov:06}\\
D8 & 0390-51900-445 & - &\citet{Izotov:06}\\
\end{longtable}
}% End longtab

\longtab[3]{
\begin{longtable}{cccccccc}
\caption{Electron densities derived from the nebular sample.}\\
\hline
\hline
Reference number& $n_e$(\sii) & $n_e$(\oii) & $n_e$(\feiii) &$n_e$(\cliii) &$n_e$(\ariv) & Adopted $n_e$  \\
& [cm$^{-3}$] & [cm$^{-3}$] & [cm$^{-3}$] & [cm$^{-3}$] & [cm$^{-3}$] & [cm$^{-3}$]  \\
\hline
\endfirsthead
\caption{Electron densities derived from the nebular sample.}\\
\hline
Reference number& $n_e$(\sii) & $n_e$(\oii) & $n_e$(\feiii) &$n_e$(\cliii) &$n_e$(\ariv) & Adopted $n_e$  \\
& [cm$^{-3}$] & [cm$^{-3}$] & [cm$^{-3}$] & [cm$^{-3}$] & [cm$^{-3}$] & [cm$^{-3}$]  \\
\hline
\endhead
\hline
\endfoot
\hline
\endlastfoot
\label{table:densities}
D1 &  -  &  -  &  -  &  -  &  -  & $100 \pm 100$ \\
D2 &  -  &  -  &  -  &  -  &  -  & $100 \pm 100$ \\
D3 &  -  &  -  &  -  &  -  &  -  & $100 \pm 100$ \\
D4 &  -  &  -  &  -  &  -  &  -  & $100 \pm 100$ \\
D5 &  -  &  -  &  -  &  -  &  -  & $100 \pm 100$ \\
D6 &  -  &  -  &  -  &  -  &  -  & $100 \pm 100$ \\
D7 &  -  &  -  &  -  &  -  &  -  & $100 \pm 100$ \\
D8 &  -  &  -  &  -  &  -  &  -  & $100 \pm 100$ \\
\end{longtable}
}% End longtab

\longtab[4]{
\begin{longtable}{cccccccc}
\caption{Measured electron temperatures from the nebular sample.}\\
\hline
\hline
Reference number& $T_e$(\nii) & $T_e$(\siii) & $T_e$(\oiii)  \\
& [K] & [K] & [K] \\
\hline
\endfirsthead
\caption{Measured electron temperatures from the nebular sample.}\\
\hline
Reference number& $T_e$(\nii) & $T_e$(\siii) & $T_e$(\oiii)  \\
& [K] & [K] & [K] \\
\hline
\endhead
\hline
\endfoot
\hline
\endlastfoot
\label{table:temperatures}
D1 &  -  &  -  & $11410 ^{+320} _{-220}$ \\
D2 &  -  &  -  & $13070 ^{+200} _{-260}$ \\
D3 &  -  & $14340 ^{+430} _{-780}$ & $14310 ^{+230} _{-150}$ \\
D4 &  -  & $13970 ^{+1160} _{-1440}$ & $14430 ^{+430} _{-360}$ \\
D5 &  -  &  -  & $13300 ^{+130} _{-160}$ \\
D6 &  -  &  -  & $12490 ^{+150} _{-290}$ \\
D7 &  -  &  -  & $11590 ^{+190} _{-230}$ \\
D8 &  -  &  -  & $11450 ^{+490} _{-530}$ \\
\end{longtable}
}% End longtab

\longtab[5]{
\begin{longtable}{cccccccc}
\caption{Adopted electron temperatures for the nebular sample. The information in this table is derived from the values in Table~\ref{table:temperatures} and the $T_e$ relations from \citet{Garnett:92} and \citet{MendezDelgado:23a}.}\\
\hline
\hline
Reference number& $T_e$(\nii) & $T_e$(\oiii) & $T_0$(O$^{2+}$)  \\
& [K] & [K] & [K] \\
\hline
\endfirsthead
\caption{Adopted electron temperatures for the nebular sample. The information in this table is derived from the values in Table~\ref{table:temperatures} and the $T_e$ relations from \citet{Garnett:92} and \citet{MendezDelgado:23a}.}\\
\hline
Reference number& $T_e$(\nii) & $T_e$(\oiii) & $T_0$(O$^{2+}$)  \\
& [K] & [K] & [K] \\
\hline
\endhead
\hline
\endfoot
\hline
\endlastfoot
\label{table:temperatures_adopted}
D1 & $10990 ^{+310} _{-210}$ & $11410 ^{+320} _{-220}$ & $9510 ^{+270} _{-180}$ \\
D2 & $12150 ^{+190} _{-240}$ & $13070 ^{+200} _{-260}$ & $10880 ^{+170} _{-220}$ \\
D3 & $13020 ^{+210} _{-140}$ & $14310 ^{+230} _{-150}$ & $11890 ^{+190} _{-130}$ \\
D4 & $13100 ^{+390} _{-330}$ & $14430 ^{+430} _{-360}$ & $11990 ^{+360} _{-300}$ \\
D5 & $12310 ^{+120} _{-150}$ & $13300 ^{+130} _{-160}$ & $11060 ^{+110} _{-140}$ \\
D6 & $11740 ^{+150} _{-270}$ & $12490 ^{+150} _{-290}$ & $10400 ^{+130} _{-240}$ \\
D7 & $11110 ^{+180} _{-220}$ & $11590 ^{+190} _{-230}$ & $9660 ^{+150} _{-190}$ \\
D8 & $11010 ^{+470} _{-510}$ & $11450 ^{+490} _{-530}$ & $9540 ^{+410} _{-440}$ \\
\end{longtable}
}% End longtab

\longtab[6]{
\begin{longtable}{cccccccc}
\caption{Ionic abundances derived from the nebular sample. All values are in 12+log(X/H) units. * indicates Fe$^{2+}$ abundance determinations based on \feiii$\lambda 4658$ with flux errors smaller than 10\%. ** indicates O$^{+}$ abundance determinations based on \oii$\lambda \lambda 7319, 7330$ auroral lines.}\\
\hline
\hline
Reference number& Fe$^{2+}$ & N$^{+}$& O$^{+}$ & O$^{2+}$ & O$^{2+}$& O$^{2+}$  \\
& & &  & \oiii-CELs $t^2=0$ & \oiii-CELs $t^2>0$ & \oiirls-RLs  \\
\hline
\endfirsthead
\caption{Ionic abundances derived from the nebular sample. All values are in 12+log(X/H) units. * indicates Fe$^{2+}$ abundance determinations based on \feiii$\lambda 4658$ with flux errors smaller than 10\%. ** indicates O$^{+}$ abundance determinations based on \oii$\lambda \lambda 7319, 7330$ auroral lines.}\\
\hline
Reference number& Fe$^{2+}$ & N$^{+}$& O$^{+}$ & O$^{2+}$ & O$^{2+}$& O$^{2+}$  \\
& & &  & \oiii-CELs $t^2=0$ & \oiii-CELs $t^2>0$ & \oiirls-RLs  \\
\hline
\endhead
\hline
\endfoot
\hline
\endlastfoot
\label{table:ionic_abundances}
D1 & $5.57 ^{+0.10} _{-0.09}$ & $6.35 \pm 0.03$ & $7.61 ^{+0.05} _{-0.04}$ & $8.04 ^{+0.04} _{-0.03}$ & $8.30 \pm 0.04$ &  -  \\
D2 & $5.29 ^{+0.12} _{-0.11}$ & $5.85 \pm 0.03$ & $7.08 ^{+0.04} _{-0.03}$ & $7.99 \pm 0.03$ & $8.23 \pm 0.03$ &  -  \\
D3 & $4.90 ^{+0.13} _{-0.12}$ & $5.63 \pm 0.02$ & $7.09 \pm 0.04$** & $7.73 \pm 0.02$ & $7.96 \pm 0.02$ &  -  \\
D4 & $5.41 ^{+0.15} _{-0.14}$ & $5.77 \pm 0.04$ & $7.25 ^{+0.08} _{-0.07}$** & $7.74 ^{+0.04} _{-0.03}$ & $7.97 \pm 0.04$ &  -  \\
D5 & $5.38 ^{+0.06} _{-0.05}$ & $6.27 \pm 0.02$ & $7.12 ^{+0.03} _{-0.02}$ & $8.00 ^{+0.02} _{-0.01}$ & $8.24 \pm 0.02$ &  -  \\
D6 & $5.39 ^{+0.07} _{-0.06}$ & $6.30 \pm 0.02$ & $7.21 \pm 0.03$ & $8.02 \pm 0.02$ & $8.26 \pm 0.03$ &  -  \\
D7 & $5.51 ^{+0.11} _{-0.09}$ & $6.10 \pm 0.02$ & $7.56 \pm 0.03$ & $8.11 \pm 0.03$ & $8.37 \pm 0.03$ &  -  \\
D8 & $5.64 \pm 0.14$ & $6.46 \pm 0.05$ & $7.64 ^{+0.09} _{-0.07}$ & $8.00 ^{+0.07} _{-0.05}$ & $8.27 ^{+0.07} _{-0.06}$ &  -  \\
\end{longtable}
}% End longtab

\longtab[7]{
\begin{longtable}{cccccccc}
\caption{Numerical values of the ICFs adopted in this work for N \citep{Amayo:21} and Fe \citep{Rodriguez:05}. These are on a logarithmic scale, so that 12+log(X/H) =  log(ICF(X))  + 12 + log(X$^{i+}$/H$^{+}$).}\\
\hline
\hline
Reference number& log(ICF(N)) ($t^2=0$) &  log(ICF(N)) ($t^2>0$)  &  log(ICF(Fe)) ($t^2=0$)  &  log(ICF(Fe)) ($t^2>0$)  \\
\hline
\endfirsthead
\caption{Numerical values of the ICFs adopted in this work for N \citep{Amayo:21} and Fe \citep{Rodriguez:05}. These are on a logarithmic scale, so that 12+log(X/H) = log(ICF(X)) + 12 + log(X$^{i+}$/H$^{+}$).}\\
\hline
Reference number& log(ICF(N)) ($t^2=0$) &  log(ICF(N)) ($t^2>0$)  &  log(ICF(Fe)) ($t^2=0$)  &  log(ICF(Fe)) ($t^2>0$)  \\
\hline
\endhead
\hline
\endfoot
\hline
\endlastfoot
\label{table:ICF_numerical}
D1 & 0.64 & 0.84  & 0.49 & 0.65  \\
D2 & 1.04 & 1.26  & 0.83 & 1.04  \\
D3 & 0.81 & 1.01  & 0.63 & 0.80  \\
D4 & 0.68 & 0.88  & 0.52 & 0.69  \\
D5 & 1.02 & 1.24  & 0.82 & 1.02  \\
D6 & 0.95 & 1.17  & 0.75 & 0.95  \\
D7 & 0.74 & 0.95  & 0.57 & 0.75  \\
D8 & 0.58 & 0.80  & 0.44 & 0.62  \\
\end{longtable}}

\longtab[8]{
\begin{longtable}{cccccccc}
\caption{Total abundances derived from the nebular sample. All values are in 12+log(X/H) units.}\\
\hline
\hline
Reference number& Fe & Fe & N & N &O & O \\
& $t^2=0$ & $t^2>0$ & $t^2=0$ & $t^2>0$  &$t^2=0$ & $t^2>0$   \\
\hline
\endfirsthead
\caption{Total abundances derived from the nebular sample. All values are in 12+log(X/H) units.}\\
\hline
Reference number& Fe & Fe & N & N &O & O \\
& $t^2=0$ & $t^2>0$ & $t^2=0$ & $t^2>0$  &$t^2=0$ & $t^2>0$   \\
\hline
\endhead
\hline
\endfoot
\hline
\endlastfoot
\label{table:total_abundances}
D1 & $6.05 \pm 0.10$ & $6.22 ^{+0.10} _{-0.09}$ & $6.99 \pm 0.03$ & $7.21 \pm 0.03$ & $8.18 \pm 0.03$ & $8.39 ^{+0.04} _{-0.03}$ \\
D2 & $6.13 ^{+0.12} _{-0.11}$ & $6.34 \pm 0.10$ & $6.90 \pm 0.03$ & $7.11 \pm 0.03$ & $8.04 \pm 0.02$ & $8.26 \pm 0.03$ \\
D3 & $5.54 ^{+0.12} _{-0.13}$ & $5.71 ^{+0.12} _{-0.10}$ & $6.46 ^{+0.05} _{-0.04}$ & $6.66 \pm 0.04$ & $7.82 \pm 0.02$ & $8.01 \pm 0.02$ \\
D4 & $5.94 ^{+0.14} _{-0.13}$ & $6.10 \pm 0.13$ & $6.46 \pm 0.04$ & $6.66 ^{+0.09} _{-0.06}$ & $7.86 \pm 0.03$ & $8.04 ^{+0.04} _{-0.03}$ \\
D5 & $6.19 ^{+0.05} _{-0.06}$ & $6.39 \pm 0.06$ & $7.29 \pm 0.02$ & $7.50 \pm 0.02$ & $8.05 \pm 0.01$ & $8.27 \pm 0.02$ \\
D6 & $6.15 ^{+0.07} _{-0.06}$ & $6.35 ^{+0.06} _{-0.07}$ & $7.26 \pm 0.02$ & $7.48 \pm 0.02$ & $8.08 \pm 0.02$ & $8.30 \pm 0.02$ \\
D7 & $6.07 ^{+0.11} _{-0.10}$ & $6.27 ^{+0.10} _{-0.09}$ & $6.83 \pm 0.02$ & $7.06 \pm 0.02$ & $8.21 \pm 0.02$ & $8.43 ^{+0.02} _{-0.03}$ \\
D8 & $6.08 ^{+0.14} _{-0.13}$ & $6.24 ^{+0.16} _{-0.14}$ & $7.05 \pm 0.05$ & $7.26 \pm 0.05$ & $8.16 \pm 0.05$ & $8.36 ^{+0.06} _{-0.05}$ \\
\end{longtable}}
% End longtab

\longtab[9]{
\begin{longtable}{cccccc}
\caption{Observed H$\alpha$/H$\beta$ fluxes prior to reddening correction in a subsample of the references presented in Table~\ref{table:IDs} where these values are reported. We also show the theoretical value calculated with PyNeb \citep{Luridiana:15}, using the atomic data from \citet{Storey:95} and the Adopted $n_e$ and $T_e$(\oiii) ($t^2$ has no influence on this RL-ratio) from Tables \ref{table:densities} and \ref{table:temperatures_adopted}, respectively.}\\
\hline
\hline
Reference number & Observed H$\alpha$/H$\beta$ & Dust-free theoretical H$\alpha$/H$\beta$ \\
\hline
\endfirsthead
\caption{Observed H$\alpha$/H$\beta$ fluxes prior to reddening correction in a subsample of the references presented in Table~\ref{table:IDs} where these values are reported. We also show the theoretical value calculated with PyNeb \citep{Luridiana:15}, using the atomic data from \citet{Storey:95} and the Adopted $n_e$ and $T_e$(\oiii) ($t^2$ has no influence on this RL-ratio) from Tables \ref{table:densities} and \ref{table:temperatures_adopted}, respectively.}\\
\hline
Reference number & Observed H$\alpha$/H$\beta$ & Dust-free theoretical H$\alpha$/H$\beta$ \\
\hline
\endhead
\hline
\endfoot
\hline
\endlastfoot
\label{table:Ha_Hb}
D28 & $3.04 \pm 0.03$ & 2.78 \\
D38 & $3.01 \pm 0.03$ & 2.79 \\
D40 & $2.85 \pm 0.03$ & 2.80 \\
D41 & $2.86 \pm 0.03$ & 2.75 \\
D53 & $2.79 \pm 0.03$ & 2.80 \\
D54 & $2.80 \pm 0.03$ & 2.78 \\
D56 & $3.07 \pm 0.03$ & 2.81 \\
D59 & $3.03 \pm 0.03$ & 2.82 \\
\end{longtable}}

\end{appendix}
\end{document}